\documentclass[journal=jacsat,manuscript=article]{achemso}

\usepackage[version=3]{mhchem} 
\usepackage{amsmath}
\usepackage{graphicx}
\usepackage{float}

\author{Mashnoon Alam Sakib}
\affiliation[UCI]
{Department of Electrical Engineering and Computer Science, University of California, Irvine, CA 92697, USA}
\altaffiliation{Contributed equally to this work}
\author{Brandon Triplett}
\affiliation[Purdue University]
{School of Electrical and Computer Engineering, Purdue University, West Lafayette, IN 47907, USA}
\alsoaffiliation{Quantum Science Center (QSC), a National Quantum Information Science Research Center of the U.S. Department of Energy (DOE), Oak Ridge, TN 37931, USA}
\altaffiliation{Contributed equally to this work}
\author{William Harris}
\affiliation[UCI]
{Department of Physics, University of California, Irvine, CA 92697, USA}
\author{Naveed Hussain}
\affiliation[UCI]
{Department of Electrical Engineering and Computer Science, University of California, Irvine, CA 92697, USA}
\author{Alexander Senichev}
\affiliation[Purdue University]
{School of Electrical and Computer Engineering, Purdue University, West Lafayette, IN 47907, USA}
\alsoaffiliation{Quantum Science Center (QSC), a National Quantum Information Science Research Center of the U.S. Department of Energy (DOE), Oak Ridge, TN 37931, USA}
\author{Melika Momenzadeh}
\affiliation[UCI]
{Department of Electrical Engineering and Computer Science, University of California, Irvine, CA 92697, USA}
\author{Joshua Bocanegra}
\affiliation[UCI]
{Department of Physics, University of California, Irvine, CA 92697, USA}
\author{Ruqian Wu}
\affiliation[UCI]
{Department of Physics, University of California, Irvine, CA 92697, USA}
\author{Alexandra Boltasseva}
\affiliation[Purdue University]
{School of Electrical and Computer Engineering, Purdue University, West Lafayette, IN 47907, USA}
\alsoaffiliation{Quantum Science Center (QSC), a National Quantum Information Science Research Center of the U.S. Department of Energy (DOE), Oak Ridge, TN 37931, USA}
\author{Vladimir M. Shalaev}
\affiliation[Purdue University]
{School of Electrical and Computer Engineering, Purdue University, West Lafayette, IN 47907, USA}
\alsoaffiliation{Quantum Science Center (QSC), a National Quantum Information Science Research Center of the U.S. Department of Energy (DOE), Oak Ridge, TN 37931, USA}
\author{Maxim R. Shcherbakov}
\email{maxim.shcherbakov@uci.edu}
\affiliation[UCI]
{Department of Electrical Engineering and Computer Science, University of California, Irvine, CA 92697, USA}
\alsoaffiliation[Second University]
{Department of Materials Science and Engineering, University of California, Irvine, CA 92697, USA}
\alsoaffiliation[UCI]{Beckman Laser Institute and Medical Clinic, University of California, Irvine, CA 92612, USA}

\title[An \textsf{achemso} demo]
  {Site-Controlled Purcell-Induced Bright Single Photon Emitters in Hexagonal Boron Nitride}

\abbreviations{hBN, hexagonal boron nitride, single photon emitter, plasmonics, strain}

\begin{document}


\newpage

\begin{abstract}



Single photon emitters (SPEs) hosted in hexagonal boron nitride (hBN) are essential elementary building blocks for enabling future on-chip quantum photonic technologies that operate at room temperature. However, fundamental challenges, such as managing non-radiative decay, competing incoherent processes, as well as engineering difficulties in achieving deterministic placement and scaling of the emitters, limit their full potential. In this work, we experimentally demonstrate large-area arrays of plasmonic nanoresonators for Purcell-induced site-controlled SPEs by engineering emitter-cavity coupling and enhancing radiative emission at room temperature. The plasmonic nanoresonator architecture consists of gold-coated silicon pillars capped by an alumina spacer layer, enabling a 10-fold local field enhancement in the emission band of native hBN defects. Confocal photoluminescence and second-order autocorrelation measurements show bright SPEs with sub-30~meV bandwidth and a saturated emission rate of more than 3.8 million counts per second. We measure a Purcell factor of 4.9, enabling average SPE lifetimes of 480~ps, a five-fold reduction as compared to emission from gold-free devices, along with an overall SPE yield of 21\%. Density functional theory calculations further reveal the beneficial role of an alumina spacer between defected hBN and gold, as an insulating layer can mitigate the electronic broadening of emission from defects proximal to gold. Our results offer arrays of bright, heterogeneously integrated quantum light sources, paving the way for robust and scalable quantum information systems.

\end{abstract}


\section{Introduction}

Single photon emitters (SPEs) are the essential building blocks for solid-state devices that are crucial for advancing quantum photonic technologies. SPEs offer vast potential to revolutionize quantum computation, communication, and sensing systems\cite{Aharonovich2009,Atature2018,Degen2017,Weber2010, o2009photonic}. To enable the practical integration of SPEs into quantum photonic integrated circuits, these nonclassical light sources must possess a combination of properties that are challenging to achieve, including a high spontaneous emission rate\cite{russell2012large, akselrod2014probing,schietinger2009plasmon}, highly radiative quantum efficiency\cite{claudon2010highly}, room-temperature operation\cite{tran2016quantum}, and deterministic emitter placement\cite{fournier2021position}.
Hexagonal boron nitride (hBN), a layered van der Waals material with a wide indirect bandgap of approximately 6~eV, is known to host a variety of SPEs that operate at room temperature, exhibiting zero-phonon line energies ranging from approximately 1.5 to 2.2~eV\cite{cassabois2016hexagonal,tran2016quantum,chejanovsky2016structural,tran2016robust,grosso2017strain1,tao2020strain2,moczala2019strain3}. The process of creating defects and optical activation usually requires techniques such as energetic ion bombardment\cite{glushkov2022engineering,pelliciari2024elementary}, irradiation and plasma engineering\cite{fischer2021controlled, xu2018single}, thermal annealing\cite{tran2016robust}, and UV-ozone treatment\cite{li2019purification}. However, these processes typically result in the random spatial arrangement of defects. Achieving site-controlled activation and large-scale integration of SPEs to quantum photonic integrated circuits is therefore an active area of research. Previous works have attempted to deterministically place the emitters using focused ion beam irradiation\cite{glushkov2022engineering,klaiss2022uncovering}, electron beam irradiation\cite{kumar2023localized}, femtosecond laser pulses\cite{gan2022large} and also by direct scratching through nanoindentation techniques\cite{xu2021creating}. An alternative approach to inducing defects by site damage is to activate and couple naturally occurring hBN defect sites with nanophotonic structures\cite{proscia2018near, ziegler2019deterministicab, mendelson2020strain, chen2024activated, IgorOriginalArrayBoronVacancy,li2021scalable}.

Efforts to improve hBN SPEs often entail designing nanoscale cavities that utilize quantum electrodynamic effects such as the Purcell effect, which spatially alters their spontaneous emission rate\cite{sauvan2013theory,goy1983observation,anger2006enhancement,wang2021cavity,vogl2019compact, svendsen2022signatures,haussler2021tunable,kim2018photonic,proscia2020microcavity,li2021scalable}. Previous works have used plasmonic nanostructures to create Purcell-enhanced emitter-cavity coupled systems\cite{tran2017deterministic,proscia2019coupling,xu2022greatly}. However, since plasmonic metals inherently suffer from scattering and non-radiative losses, optical field-enhancing nanoplasmonic geometries require stringent design and fabrication criteria to be met for achieving an effective coupling to an emitter. Some of these requirements include spatial alignment of an emitter to the resonant cavity electric fields and also an optimized gap between the emitter and nearby metallic surface for avoiding quenching of fluorescence\cite{kongsuwan2018suppressed, anger2006enhancement}. So, an efficient plasmonic nanocavity-integrated antenna architecture using radiative decay engineering approaches could have immense potential to enhance the emitter-cavity mode coupling mechanism and may have far-reaching implications for both fundamental and applied research. 

In this work, we achieve site-controlled creation of optically activated, bright SPEs in a plasmonic nanoresonator (PNR) antennas draped with unprocessed few-layer hBN. The gold-coated silicon nanopillars act as antenna nanoresonators supporting broadband surface plasmon resonances that create plasmonic hot-spots with a strong E-field confinement. The emitters, placed into near-contact with the gold nanoparticles, experience a dramatically enhanced spontaneous emission rate 
\cite{sauvan2013theory,goy1983observation,maksymov2010metal,derom2012resonance,anger2006enhancement,kuhn2006enhancement,galloway2009ultrafast}. 
We experimentally measured an average Purcell factor of 4.9 with the incorporation of alumina-coated plasmonic nanoresonators that result in shortened SPE lifetimes by a factor of five with an average of 480~ps, compared to 2.3~ns from non-plasmonic arrays. Significant radiative enhancement of photoluminescence (PL) resulted in bright room-temperature SPEs with sharp emission bandwidth down to sub-30 meV along with a saturated photon count rate of more than 3.8 million counts/s. We also observed an overall SPE yield determined by zero-delay second-order autocorrelation $g^{(2)}(0)<0.5$ of approximately 21\% over 110 measured sites. We performed density functional theory (DFT) calculations on SPEs hosted within hBN layers placed on a gold substrate. These calculations, supported by the experimental PL and $g^{(2)}(\tau)$ measurements, reveal electronic and electromagnetic quenching for intrinsic defects in close proximity to gold. 
This work demonstrates a viable path towards realizing scalable, room-temperature quantum photonic devices utilizing deterministically placed, Purcell-enhanced SPEs in hBN.



\section{Results and discussion}

\subsection{PNR-induced hBN SPEs}

\begin{figure}[ht!]
\centering
\includegraphics[width=0.9\textwidth]{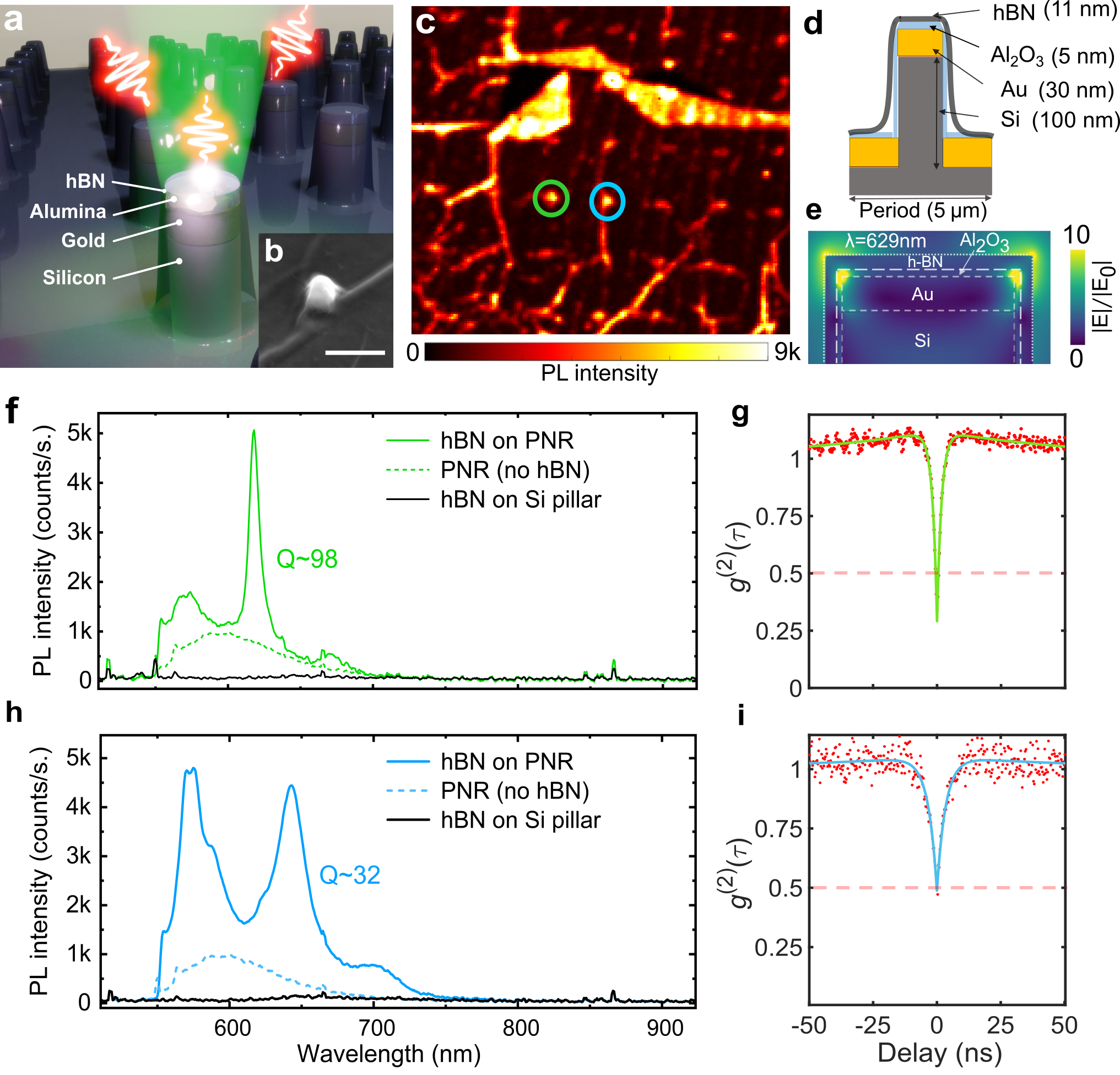}
\caption{PNR-induced SPEs in hBN at room temperature. (a) Schematic of multilayer hBN coupled to a PNR. The hBN is separated from the plasmonic cavity by a 5 nm alumina spacer layer. The wavefunction in green and red indicates the Purcell-enhanced SPE absorbing and then emitting a photon. (b) Scanning electron microscopy image of a single PNR draped around with multilayer hBN. Scale bar is 500 nm. (c) A confocal scanning photoluminescence (PL) intensity map of a 6 x 7 emitter array containing hBN draped PNRs. Scale bar is 2 $\mu$m. (d) Schematic cross-section of the hBN draped PNR (e) Cross-sectional view of the simulated electric field enhancement $|E|/|E_0|$ distribution profile under normal illumination. The E-field distribution illustrates the confinement of the plasmonic hotspots at the resonator perimeter containing the draped hBN. (f) PL spectra and (g) corresponding $g^{(2)}(\tau)$ from the emitter recorded from the green circled regions of the PL intensity mapping from (d). (h) PL spectra and (i) corresponding $g^{(2)}(\tau)$ from the emitter recorded from the blue circled regions of the PL intensity mapping from (d).}
\label{fig1}
\end{figure} 

PNR consists of a 30-nm-thick gold and a 5-nm-thick alumina spacer layer deposited on top of silicon pillar arrays with a height of 100 nm and a diameter of 180 nm. First, large-scale pillar arrays are fabricated in an intrinsic silicon substrate using electron beam lithography. The pillars are coated with gold by e-beam evaporation and alumina by atomic layer deposition. Following this, a multilayer 11-nm-thick hBN film is wet-transferred onto this fabricated PNR array (see Methods and Supplementary Section S1). The hBN receives no annealing, radiation, or other processes commonly used to create emitters. Upon the wet transfer, the multilayer hBN drapes around the PNR in a “tent-pole” like topographical profile (Figure 1a). Figure 1b shows a scanning electron microscope image of a representative PNR draped with multilayer hBN. Figure 1c represents a confocal scanning PL intensity map of the final assembled structure recorded at room temperature (see Methods). Here, the spots with point-like emission marked with green and blue circles correspond to the regions where the transferred hBN is draping around the PNR sites and consequently inducing SPEs. We can see the detailed structural parameters along with a cross-sectional geometry illustrated schematically in Figure 1d.  In Figure 1e, the fundamental plasmonic mode of a multilayer hBN coupled to alumina-coated PNR is numerically calculated by the finite-difference time-domain (FDTD) method (see Methods and Supplementary Section S2). The simulated electric field distribution shows that at the resonance wavelength of 629 nm, the deep sub-wavelength gold films stimulate the generation of plasmons in volumes much smaller than the wavelength and gives rise to a 10-fold enhanced plasmonic hot-spots along the gold edges. This resonance wavelength is primarily determined by the size and design of the sub-wavelength plasmonic nanoantenna and its' environment\cite{akselrod2014probing,lassiter2013plasmonic}. When hBN film is draped on PNRs, the emitters could potentially be placed in the vicinity of these PNR supported hot-spots. When the emitters have a spectral and spatial overlap with the plasmonic modes, near-field enhancement takes place\cite{sauvan2013theory,maksymov2010metal}. According to the Fermi's golden rule, in the weak-coupling regime, this potentially helps in establishing an effective out-coupling between the emitter and the cavity and considerably increases the spontaneous emission rate of the systems\cite{sauvan2013theory}. With the alumina spacer layer additionally helping in suppressing the emitter quenching\cite{kongsuwan2018suppressed,anger2006enhancement}, an enhanced Purcell-effect could be realized through the PNRs that effectively leads to the generation of SPEs.

The spectra in Figure 1f represent PL emission recorded from the resonator site marked with a green circle in Figure 1d. The spectra show a sharp zero-phonon line (ZPL) at 618 nm (2.01 eV) with a full width at half maximum (FWHM) of ~20 meV. This ZPL, with an extracted quality factor (Q) of 98 (see Supplementary Section S3), was accompanied by a 158 meV redshifted phonon sideband (PSB) at 671 nm (1.85 eV). Second-order autocorrelation, $g^{(2)}(\tau)$ measurements were performed at room temperature with a Hanbury-Brown and Twiss (HBT) setup to record the nonclassical nature of single photon emission from this PNR-coupled emitters. Figure 1g shows that the antibunching dip is at $g^{(2)}(0)$ = 0.29 (± 0.02), proving that this hBN defect site was behaving as an SPE. Similarly, the PL spectra from the nearby emission site (blue circle in Figure 1d), are shown in Figure 1h. This shows two ZPL lines appearing at 576 nm ($\mathrm{ZPL_1}$ at 2.15 eV) and 643 nm ($\mathrm{ZPL_2}$ at 1.93 eV), respectively. $\mathrm{ZPL_2}$ had a PSB 162 meV situated at 702 nm (1.76 eV). In both Figure 1f and 1h, this difference in energy gap between ZPL and PSB lay within the range of 160 ± 5 meV as reported in previous works for the case of multilayer hBN defects\cite{tran2016robust}. A detailed analysis of representative SPE PL spectra has been provided in the Supplementary Information Sections S4, S5 and S6. Variation in respective ZPL positions from one PNR to another may be attributed to the local variations in the mechanical draping profile of the hBN along the different resonators. The corresponding antibunching measurement shown in Figure 1i represents its single photon emission behavior with a recorded $g^{(2)}(0)$ = 0.47 (± 0.04). Here, in Figures 1f and 1h, the SPE PL spectra (green and blue solid lines) show an average integrated PL enhancement by a factor of 7.8 and 14, respectively, when compared to the PL recorded from hBN draped on bare silicon pillars (black solid line). These enhancements are likely to originate from the overlap among the photon emitters and the plasmonic resonance supported by the PNR antennas. For the case of hBN on bare Si pillars, due to the absence of plasmonic resonance modes, the defect sites on Si pillars do not show any quantum emission (see Supplementary Section S7 for the different geometries and height ranges of nanoresonators that were studied).

Our hBN film did not undergo any pre-processing proposed previously for defect activation \cite{tran2017deterministic, fischer2021controlled,zhong2024carbon,chejanovsky2016structural,grosso2017tunable}. As a result, the commercially available hBN film grown by chemical-vapor deposition (CVD) hosts only the randomly distributed intrinsic defects that naturally occur during the growth process\cite{wong2015characterization,tran2016quantum}. In this experiment, SPE behavior is only found for the case of hBN on PNRs (shown in Figures 1g and 1i), as the Si-only pillar sites were not found to support any SPEs. This indicates that strain induced by Si pillars may not be enough to activate SPEs from the intrinsic hBN defects on sites. This suggested that the naturally occurring hBN defects must be coupled to the PNRs to enable an effective coupling mechanism along with any strain-mediated effects. In previously reported works, the strain has been shown to have a role in the modification of local bandstructure in hBN for SPE activation\cite {mendelson2020strain,proscia2018near,chen2024activated}. We note that in our case, the synergistic effect of nanoscale strain-perturbation and plasmonic mode coupling may be playing an important role in achieving a Purcell-enhanced SPE phenomenon.




\begin{figure} [h]
\centering
\includegraphics[width=1\textwidth]{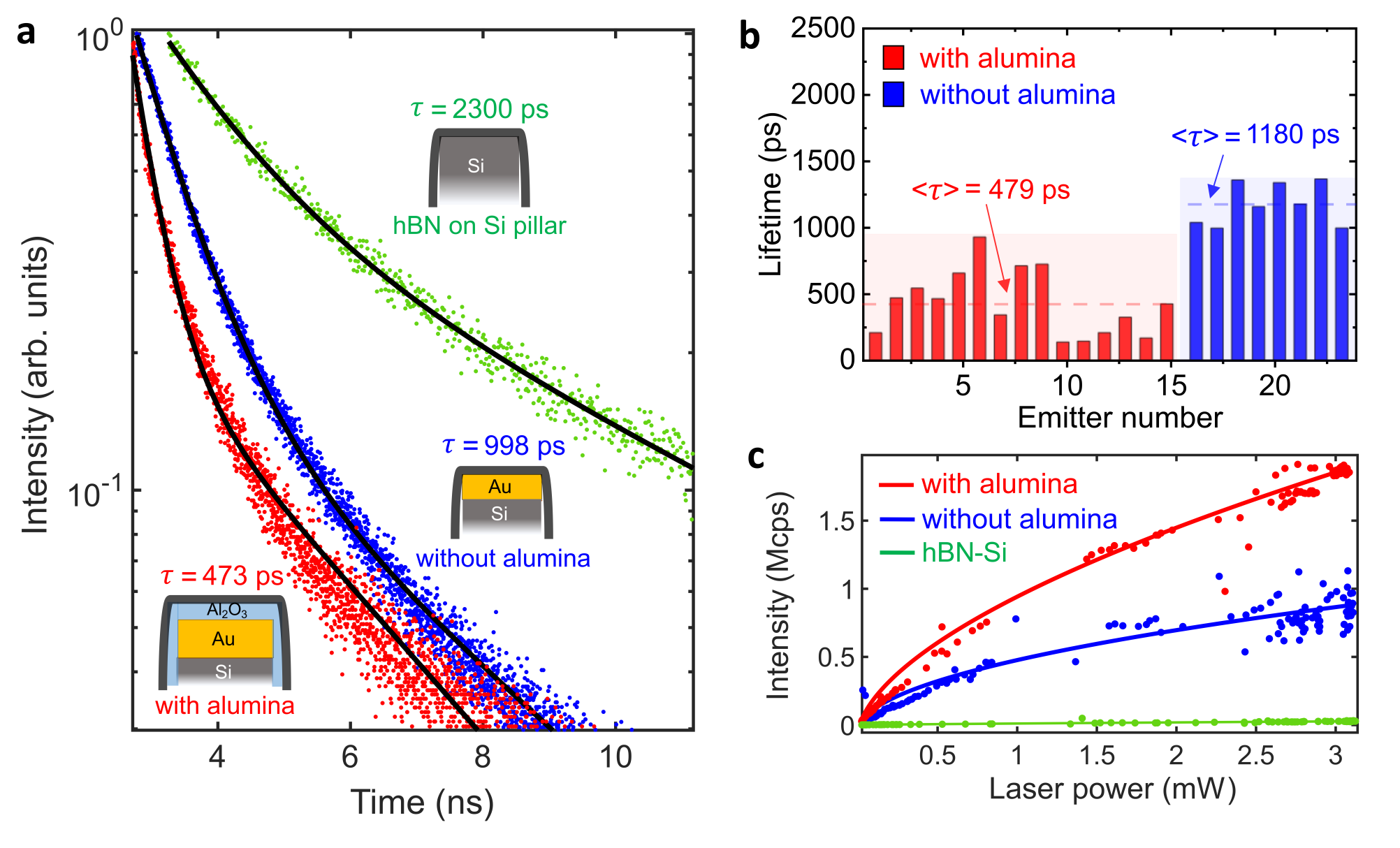}
\caption{Purcell-enhanced photophysical characteristics of SPEs in hBN. (a) Time-resolved PL decay showing radiative transition lifetimes recorded from three different device configurations with hBN draped on alumina coated PNR (red dot), PNR without alumina (blue dot), and bare silicon resonator (green dot). Solid lines represent corresponding fits to a bi-exponential decay function model. (b) Histogram showing the statistical distribution of radiative transition lifetimes recorded from time-resolved photoluminescence measurements of 23 photon emitters recorded from hBN on alumina-coated PNR (red box) and PNR without alumina (blue box) (c) Fluorescence saturation measurements recorded from three different device configurations mentioned in (a). Corresponding fits in solid lines are plotted using a first-order saturation model.}
\label{fig2}
\end{figure}

\subsection{Photo-physical Characteristics of Purcell-enhanced SPEs}

To quantify the Purcell-enhanced single photon emission, we compare the spontaneous emission rate observed from a PNR-coupled emitter with the non-plasmonic case. This measured rate enhancement directly corresponds to the cavity Purcell factor ($F_p$)\cite{sauvan2013theory,akselrod2014probing}. The Purcell factor can be directly expressed as the ratio of the emitter transition rate ($\Gamma$) with the hBN on PNRs to a native emitter transition rate ($\Gamma_0$) with the hBN on bare Si pillars. Here, the transition rate is the sum of all the radiative and nonradiative transition rates. So, $F_p$ can be also be defined as the inverse ratio of the lifetime of the emitter ($\tau$) with hBN on PNRs to the native emitter lifetime ($\tau_0$) with hBN on bare Si pillars:

\begin{equation}
F_p = \frac{\Gamma}{\Gamma_0} = 
\frac{
\Gamma_{r} + \Gamma_{nr}}
{\Gamma_0} = \frac{\tau_0}{\tau},
\end{equation}
where $\Gamma_{r}$ and $\Gamma_{nr}$ are radiative and non-radiative rates, respectively. To demonstrate the emission rate control, in Figure 2a, we show the picosecond laser-assisted time-resolved PL measurements. We measured across three samples with hBN draped over (1) bare silicon pillars with no plasmonic coating (green dots), (2) PNRs with no alumina coating (blue dots), and (3) PNRs with a 5-nm-thick alumina spacer coating (red dots), respectively. We measured the time-resolved photoluminescence using a 515 nm pulsed laser with an 80 MHz repetition rate and 40 ps pulse width. We fitted these measured fluorescence decays to a two-level bi-exponential model function and retrieved the optical transition lifetimes. Here, the recorded emission lifetime for the case of hBN on bare silicon pillars reaches 2.3 ns. In the case of hBN on PNRs with alumina coating (red dots) and without any alumina coating (blue dots), the lifetimes get reduced to 998 ps and 473 ps, respectively. This reduction in lifetime suggests that the spontaneous emission rate is getting enhanced by a Purcell factor of 2.3 and 4.9, respectively. This supports the hypothesis that the presence of an alumina spacer layer allows for higher Purcell enhancement by suppressing quenching. To statistically probe this Purcell-enhancement dependence on the presence of a spacer layer, we also measured this fluorescence lifetime across 23 different emitters both in the presence and in the absence of an alumina layer. In Figure 2b, we show the statistical analysis of these recorded emitters with a histogram. Here, without alumina, the recorded average emission lifetime was ~1180 ps. For alumina-coated PNRs, the average lifetime was ~480 ps, with a 2.5-fold reduction. 

To gain further insights into the photophysical properties of the emitters, we characterized fluorescence saturation photon count rates by measuring the laser-excitation-power-dependent PL intensity across the three cases shown in Figure 2a by measuring the PL intensity as a function of the excitation power. These results are shown in Figure 2c. To extract the saturated single-photon count rates, we corrected the measured datasets for the background emission and fitted them using the first-order saturation model.  
\begin{equation}
I(P) = \frac{I_{\infty} P}{P_{sat}+P},
\end{equation}
where $I_{\infty}$ and $P_{sat}$ are fitting parameters corresponding to the saturated emission rate and saturation power, respectively\cite{senichev2021room,fournier2021position,tran2016robust}. In the case of hBN situated on PNR coated with alumina (red) and without alumina (blue), the saturated emission rates were recorded as 3.8~Mcps (with $P_{sat}$ = 3.2 mW) and 1.3~Mcps (with $P_{sat} = 1.8$~mW), respectively. For hBN situated on bare Si pillars (green), the $I_{\infty}$ was measured as 0.18~Mcps (with $I_{\infty} = 1.7$~mW) (see Supplementary Section S8 for the maximum saturated emission rate of 7.9~Mcps). These measurements showed overall radiative enhancement factors of 21.1 for hBN on alumina-coated PNR and 7.2 for hBN on PNR without alumina coating, compared to the hBN on bare Si pillars. This suggests that the overall radiative enhancement factors for saturation enhancements are comparable to the maximum Purcell factors achieved through radiative lifetime measurements mentioned in Figures 2a and 2b. Overall, these photophysical characteristic measurements shown in Figure 2 suggest that in the case of hBN coupled to alumina-coated PNRs, the efficiently localized electric-field enhancement and suppressed non-radiative scattering leakage allow the Purcell-enhanced cavity dynamics to play a major role in accelerating the spontaneous emission decay rate towards realizing fast and bright SPEs. 




\subsection{Alumina-suppressed Emitter Quenching}

\begin{figure}
\centering
\includegraphics[width=1\textwidth]{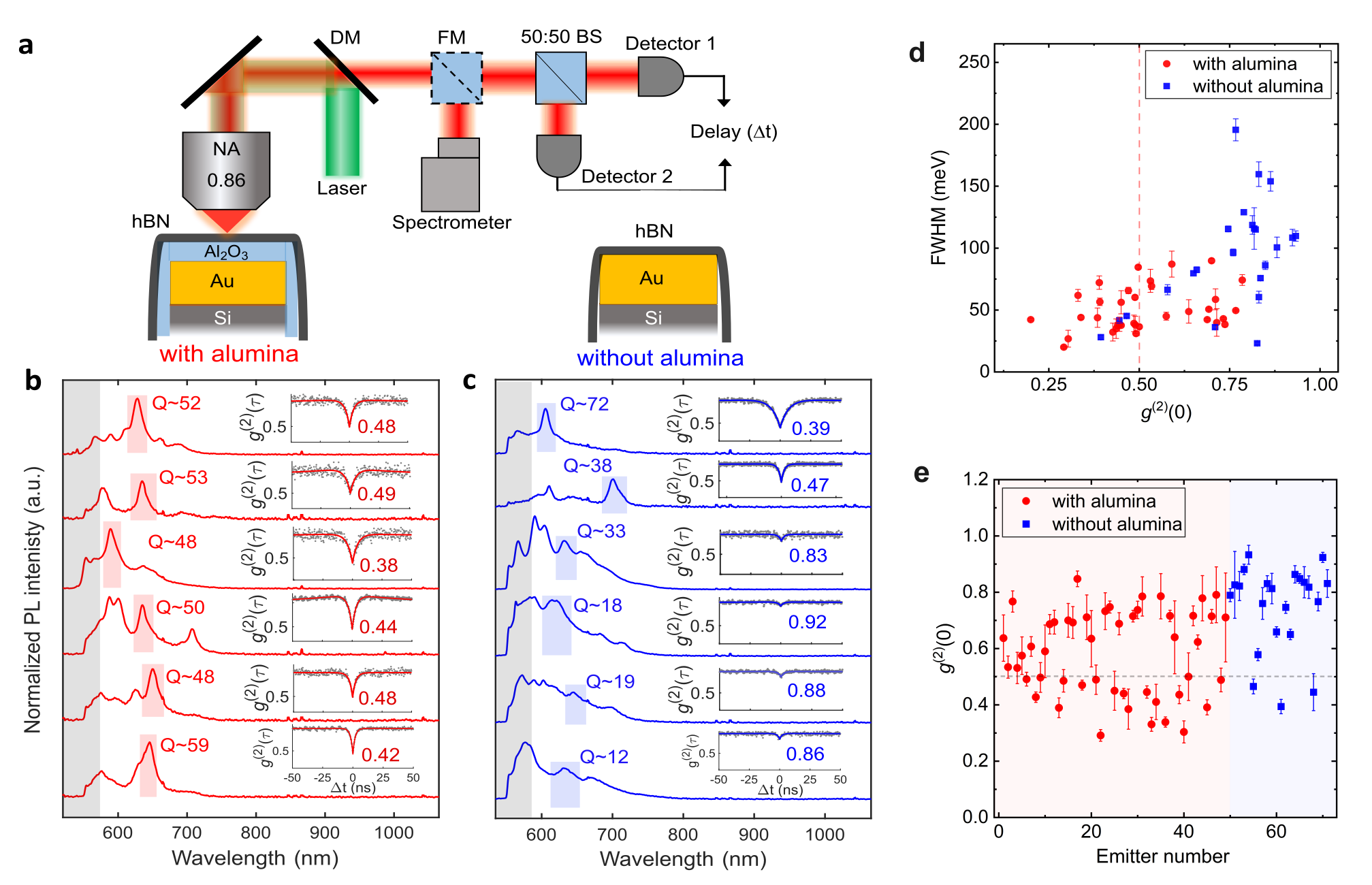}
\caption{Optical characterization of cavity-enhanced single photon emitters created by the plasmonic nanoresonators. (a) Simplified schematic of the confocal PL setup integrated with a Hanbury-Brown and Twiss (HBT) measurement scheme that was used to record PL intensity and second-order autocorrelation $g^{(2)}(\tau)$ measurements at room temperature. The objective lens, reflecting mirror, dichroic mirror, flippable mirror, beam splitter, and single avalanche photon detectors are denoted by Obj., DM, FM, BS, and Detector, respectively. (b and c) Normalized PL spectra of six representative photon emitters and (insets) corresponding $g^{(2)}(\tau)$ measurements recorded from hBN on PNR with alumina (b) and without alumina (c). The spectra are normalized without background correction and offset vertically. The solid traces in the insets represent theoretical fits to the recorded $g^{(2)}(\tau)$ data. (d) The FWHM and the antibunching dip $g^{(2)}(0)$ values extracted from measurements recorded for 58 PNR-coupled emitters. (e) Scatter plot of the $g^{(2)}(0)$ values extracted from a total of 72 emitters from hBN on PNR (red dot) with alumina and (blue dot) without alumina device configurations. The error bars represent their corresponding fitting uncertainties. Background subtraction has not been applied to any recorded autocorrelation data.}
\label{fig3}
\end{figure}



To gain insights into the cavity-enhanced SPE performance, we optically characterized more than 70 emitters and analyzed their measured PL responses and corresponding second-order autocorrelation $g^{(2)}(\tau)$ measurements at room temperature. Details of the schematic and apparatuses used for the custom-built scanning confocal microscopy setup can be found in Methods (Optical Characterization) and Supplementary Information Figure S9. We measured two device configurations, one in which the hBN film is coupled to PNRs with an alumina spacer layer and one without the alumina spacer layer, as shown in Figure 3a. Figures 3b and 3c show examples of recorded PL spectra and their corresponding $g^{(2)}(\tau)$ measurements for PNRs having an alumina spacer layer and without an alumina spacer layer, respectively. In Figure 3b, PL spectra show that quantum emission mostly occurred within a spectral bandwidth of 90 nm that spans from 585 nm (2.11 eV) to 675 nm (1.83 eV). These spectra also show that emitters coupled to the alumina-spacer layer, on average, had higher Q-factors with lower emission full-width at half maximum (FWHM) as compared to the latter case of PNRs without alumina. The alumina coating acts as a deep-subwavelength dielectric spacer, preventing electromagnetic quenching and homogenizing the local density of photonic states. This effect has been theoretically predicted in the case of deep-subwavelength nanoresonators\cite{sauvan2013theory, maksymov2010metal,goy1983observation,derom2012resonance}. When the hBN was in direct contact with the gold without the spacer layer, the recorded PL spectra (Figure 3c) mostly consisted of multiple spectral peaks that had lower relative intensities as compared to the case with the spacer layer. This increase in emitter PL is likely due to an acceleration of the radiative decay rate of the coupled emitters and allows for an overall higher yield of SPE generation across the cites\cite{kongsuwan2018suppressed,sauvan2013theory,anger2006enhancement,kuhn2006enhancement, galloway2009ultrafast}. In Figure 3d, the values of the emission FWHM and their corresponding antibunching $g^{(2)}(0)$ values are plotted for 58 representative photon emitters recorded from hBN draped on PNRs with alumina spacer-layer coating (red dots) and without alumina (blue squares). In the presence of alumina, the emitters achieved an overall average sub-50 meV emission with 21 out of the total 35 emitters behaving as SPEs with $g^{(2)}(0)<0.5$. For the case without alumina, the emission becomes broader with an average FWHM of over 90 meV and only 3 out of 21 emitters are of SPE nature. Moreover, Figure 3e shows the overall quantum yield of our fabricated devices for the cases of with alumina (red dots) and without alumina (blue dots). The emitters showing $g^{(2)}(0)$ below the threshold value of 0.5 confirms their SPE behavior. Based on the measurement of 110 emitters, we recorded 24 SPEs across the two device configurations with an approximate yield of 21\%. Figure 3e shows 72 representative $g^{(2)}(\tau)$ measurements recorded across the 110 emitters. It also supports our claim that the presence of a 5-nm-thick alumina spacer layer helps in achieving a comparatively higher success rate of SPE formation.





\subsection{DFT Calculations of Defect State Spectra}

We note that SPEs from samples without an alumina spacer exhibit broader average linewidths (Figure 3d) and a degraded $g^{(2)}(0)$ response (Figure 3e). One reason for broader linewidths in PNRs without an alumina spacer could be the coupling to the image charge in gold, which leads to damped single-photon transitions. However, the time-bandwidth product of the single photon peaks significantly exceeds unity, indicating that peak widths are not primarily determined by transition lifetimes. Thus, image charge induced damping does not fully explain the variation in linewidths. To better understand the issue of linewidths and SPE purity, we turn to density functional theory (DFT) with a model system of the hBN/Au interface. DFT calculations are performed for a system of three-layer hBN supported on an Au substrate. As mentioned above, the hBN film used in our work did not undergo any established defect-generation processes. Since the commercially bought hBN film had been CVD grown, the native defect sites are likely to be intrinsic\cite{wong2015characterization,tran2016quantum}. Moreover, the ZPL and PSB wavelength distribution (see Supplementary Section S6 and S10) hints that the most likely candidate could be (1) nitrogen vacancy $\mathrm{V_N}$ and (2) a complex anti-site where the nitrogen replaces the boron site with a missing atom at the nitrogen site $\mathrm{N_BV_N}$\cite{tran2016robust,tran2016quantum}. Therefore, we focused our DFT calculations on both these $\mathrm{V_N}$ and $\mathrm{N_BV_N}$ defects, with results separately obtained for defects on each layer to characterize proximity effects.

\begin{figure}
\centering
\includegraphics[width=1\textwidth]{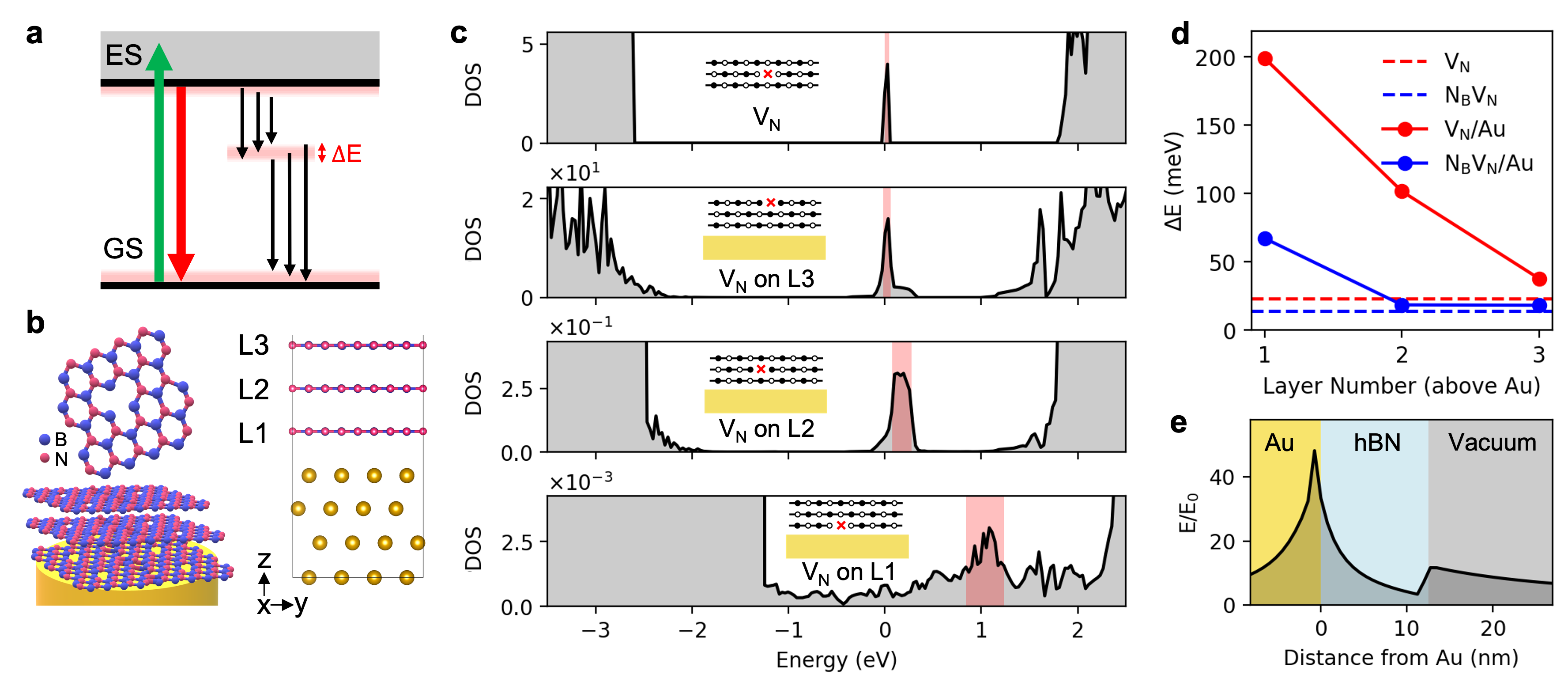}
\caption{(a) Energy diagram illustrating the effect of broadening the mid-gap state, ground state (GS), and excited state (ES) on the photoluminescence spectrum. (b) Depiction of defect structure $\mathrm{N_V}$ (bottom left) within hBN as analyzed through DFT calculations. Defect structure $\mathrm{N_V}$ is considered at three layer positions ($\mathrm{L_1}$, $\mathrm{L_2}$, $\mathrm{L_3}$) above a gold (Au) slab. (bottom right) (c) Partial density of states (DOS) of the three-layer hBN system in four configurations: with $\mathrm{N_V}$ on the middle layer and no Au slab (top panel) and $\mathrm{N_V}$ on each of the three layers in proximity to Au (bottom three panels). Full width at half maximum (FWHM) of the mid-gap states shaded in red shows significant broadening from the interaction with Au. (d) FWHM of the $\mathrm{V_N}$ (red) and $\mathrm{N_BV_N}$ (blue) mid-gap states as a function of layer number from the Au surface. Dashed lines correspond to defect structures in the absence of Au. (e) Plasmonic enhancement of electric field versus distance from Au, obtained from Lumerical FDTD results. The data presented corresponds to a horizontal cut positioned in the vicinity of the maximum field enhancement.}
\label{fig3}
\end{figure}

DFT results indicate significant quenching and broadening of the mid-gap states associated with vacancies near gold (Figure 4c). This broadening, along with the observed softening of the conduction and valence band edges, facilitates numerous nearly degenerate single-photon transitions, illustrated in Figure 4a. These effects could reduce the purity of the single photon source and lead to broader spectral peaks. While the gold’s influence quickly diminishes with layer depth (Figure 4d), the electric field is found to strongly localize near the hBN/Au interface (Figure 4e) where electronic effects are likely still relevant. In contrast, an alumina spacer leads to a more evenly distributed field across hBN (see Supplementary Figure S11e), suggesting that the observed defect states are less likely to be positioned near an interface. We further observe negligible broadening when hBN is supported on alumina, as illustrated in Supplementary Figure S12d. These findings help explain the differences in the character of SPEs with and without an alumina spacer and support the beneficial role of an insulating layer between hBN and Au.

\section{Conclusions}

In summary, we have demonstrated a PNR antenna architecture for the experimental realization of deterministically placed Purcell-enhanced SPEs in a hBN at room temperature. We have demonstrated activation and enhancement of hBN SPEs by utilizing the coupling of 
natural hBN defects with the resonance modes of the PNR platform, accelerating the coupled emitters' spontaneous emission rate. 
An average of 5-fold reduction in lifetime, down to an average of 480~ps, is observed for bright SPEs with a saturated photon count rate of up to $3.8 \times 10^6$~counts/s and a yield among all measured sites of 21\%. We also find that suppression of line broadening by use of an alumina spacer layer is favorable to the performance and yield of the emitters. DFT calculations further emphasize the beneficial role of the alumina spacer for intrinsic V$_N$ and N$_B$V$_N$ defects in hBN by SPE linewidth broadening mitigation. Our findings highlight the potential of natural defects in unprocessed hBN on resonant nanostructures as a versatile platform for site-controlled SPEs, offering stable operation at room temperature and emission across a broad spectral range. 
Our results represent an important step toward scalable room-temperature light-based quantum information systems.

\section{Methods and Materials}

\subsection{Alumina-coated Plasmonic Nanoresonator Array Fabrication}

Arrays of silicon pillars are fabricated using standard techniques of electron beam lithography (EBL) and reactive ion etcher (RIE). First, a negative electron-beam resist, hydrogen silsesquioxane (HSQ, Dow Corning), was spin-coated on Si wafer at 100 rpm for 10 s followed by another step with 5000 rpm for 60 s. The coated silicon wafer was then baked at 120 degrees C for 2 min and at 180$^{\circ}$ C for another 2 min. Electron beam lithography (JEOL JBX 6300-FS) was used with a dose of 300~$\mu$C/cm\textsuperscript{2} to transfer the target patterns. Following the patterning, the sample was developed using a solution made from a mixture of 1 wt,\% NaOH, and 4 wt,\% NaCl in DI water for 4 min and then rinsed with DI water. The HSQ patterns were transferred to silicon by using an inductively coupled reactive ion etcher (Oxford Plasmalab 100) with a mixture of sulfur hexafluoride (SF\textsubscript{6}) and nitrogen (N\textsubscript{2}) gas. Following the pattern transfer, the whole sample was immersed in a resist etchant solution to remove the resist layers. The sample is then rinsed with DI water and blow-dried with nitrogen. To coat plasmonic material on top of the fabricated silicon resonator arrays, electron beam evaporation (Temescal CV6) was used to deposit 1~nm of chromium (Cr) as an adhesion layer, followed by 30~nm of gold (Au). During this step, the gold deposition rate was carefully maintained at 1~Å/s to ensure a uniform and smooth surface. Following the Au deposition, atomic layer deposition (ALD, Machine, Model) was used to deposit 5~nm of alumina.

\subsection{hBN Wet-transfer Protocol}

Multilayer hBN samples on copper (Graphene Supermarket) had an average thickness approximately 11~nm (30 layers). The hBN as grown by chemical vapor deposition on both sides of a 20~$\mu$m thick copper foil. The hBN on one side was etched away with oxygen plasma (50 sccm of O\textsubscript{2} at 150 mTorr for 4 min at 150 W). The remaining hBN was spin-coated with 100 $\mu$L of 4\% poly(methyl methacrylate) in anisole (PMMA A4) at 100 rpm for 10 seconds followed by a 3000 rpm for 1 min. The PMMA-coated sample was soft-baked on a hotplate at 90$^{\circ}$ C for 90 s. The PMMA-coated hBN sample was then carefully floated on a Cu etchant (0.1~M ammonium persulfate solution). After 10-12 hours, when the copper is completely etched, the PMMA-coated hBN becomes transparent and keeps floating in the solution. It is then carefully scooped with a glass side and carefully transferred into a diluted HCl solution, removing the copper etchant remnants. After 3 DI water washes, the PMMA-coated hBN is then carefully scooped with the pre-cleaned silicon nanoresonator target substrate and put on the hotplate at 100$^{\circ}$ C for approximately 10 min. This allows the hBN to get a good adhesion onto the substrate. The PMMA-coated hBN on the target substrate is submerged in room-temperature acetone for approximately 12 hours. For the final step, the hBN on the target substrate is exposed to acetone vapor for 1 hour, uniformly dissolving the remnant PMMA. The sample is then left to dry at room temperature overnight.

\subsection{Optical Characterization}

Optical characterization was conducted using a custom-built scanning confocal microscopy setup, constructed around a Nikon Ti-U inverted microscope body. The PL scanning was performed by scanning a Nikon LU Plan 100x/0.90 NA objective with a P-561 PIMars piezo stage upon which it was mounted, which is controlled by an E-712 controller and Alignment Firmware provided by Physik Instrumente. Continuous wave excitation experiments were facilitated by a 200 mW 532 nm laser sourced from RGB Photonics. [Insert laser Spot size] To isolate the desired fluorescence from the excitation source, a main dichroic beamsplitter (550 nm long-pass DMLP550L, Thorlabs) was employed, followed by 550 nm and 575 nm long-pass filters (FEL0550, Thorlabs). The collected fluorescence was passed through a 100 $\mu$m pinhole to spatially filter the light. For auto-correlation measurements, two single-photon avalanche detectors (SPADs) were utilized, having a 30 ps time resolution and 35\% quantum efficiency at 650 nm (PDM, Micro-Photon Devices). Scanning and saturation measurements were carried out using a SPAD with 69\% quantum efficiency at 650 nm (SPCM-AQRH, Excelitas). Spectral analysis was facilitated by a QE65000 visible-to-near infrared spectrometer from Ocean Insight.

\subsection{DFT calculations}

DFT calculations were performed using the Vienna Ab initio Simulation Package (VASP) with the PBE pseudopotential. The van der Waals interactions were treated with the DFT-D3 correction\cite{grimme2011effect}. To address lattice mismatches for heterostructures, substrates were strained by -2.5\% for Au and 3.5\% for $\mathrm{Al_2O_3}$. The geometries were relaxed until the force on each atom was below 0.01 eV/Å. In the heterostructures, all atoms except those in the upper Au layer were fixed, whereas all hBN atoms were free to move. To reduce periodic interactions between defects, a 4x4x1 supercell of hBN was utilized for the alumina heterostructure and the hBN trilayer without a substrate. We consider only Al-terminated (Al-I) alumina, which has previously been demonstrated as the most energetically favorable\cite{kurita2010atomic}. hBN was placed over  $\mathrm{Al_2O_3}$  with a horizontal alignment that positions an N atom of layer-1 over an Al atom, as was found to be most stable. 

A $\Gamma$-centered 4x4x1 k-mesh is used for these calculations. For hBN on Au, a different primitive cell of hBN was employed to align better with the Au lattice, with a 4x2x1 cell proving adequate. Here, the k-mesh is set to 4x3x1. In all cases an energy cutoff of 450 eV is used and electronic convergence is set to $10^{-5}$ for geometry relaxation and $10^{-6}$ for self consistency calculations. A vacuum region of more than 15 Å is included in all cases to reduce periodic interactions. 

The electronic bandgap of hBN is known to be underestimated with DFT and is often improved using hybrid functionals such as the Heyd-Scuseria-Ernzerhof (HSE) functional. However, our findings are intended to be qualitative and do not depend on the numerical accuracy of the hBN bandgap. Moreover, the large system sizes in our work make such corrections prohibitively expensive.

\subsection{FDTD Simulations}

Finite-difference time-domain (FDTD) simulations are performed with Ansys Lumerical FDTD package to calculate the scattering spectra and the associated electric field distributions at the hBN-PNR architecture. A total-field scattered-field (TFSF) formulation is used to excite the hBN-PNR structures. The incident plane wave injected from the TFSF source is transverse-magnetic polarized and arrives upon the interface of hBN-draped PNR under normal incidence. The 3D simulation region is calculated with a uniform mesh size of 0.5 nm in the x-, y- and z- directions. Tabulated data were used to model the optical properties of silicon and alumina\cite{palik1998handbook}, gold\cite{johnson1972optical}, and hBN\cite{lee2019refractive}. Frequency-domain field and power monitors are placed that extended along the interfaces for recording the near-field E-field intensity enhancements across on- and off-resonance peak points. Perfectly matched layers are added at the boundaries to represent an isolated PNR.

\begin{acknowledgement}

This work is supported by the National Science Foundation (ECCS-2339271), Defense Advanced Research Projects Agency (D22AP00153), U.S. Department of Energy (DOE), Office of Science through the Quantum Science Center (QSC), DE-AC05-00OR22725 and the Air Force Office of Scientific Research (AFOSR) grant FA9550-22-1-0372.

\end{acknowledgement}


\providecommand{\latin}[1]{#1}
\makeatletter
\providecommand{\doi}
  {\begingroup\let\do\@makeother\dospecials
  \catcode`\{=1 \catcode`\}=2 \doi@aux}
\providecommand{\doi@aux}[1]{\endgroup\texttt{#1}}
\makeatother
\providecommand*\mcitethebibliography{\thebibliography}
\csname @ifundefined\endcsname{endmcitethebibliography}
  {\let\endmcitethebibliography\endthebibliography}{}

\newpage
\begin{suppinfo}
\renewcommand{\thefigure}{S\arabic{figure}}
\setcounter{figure}{0}

\subsection{Section S1. hBN Wet Transfer and Raman Measurements}
\begin{figure}[h]
\centering
\includegraphics[width=1\textwidth]{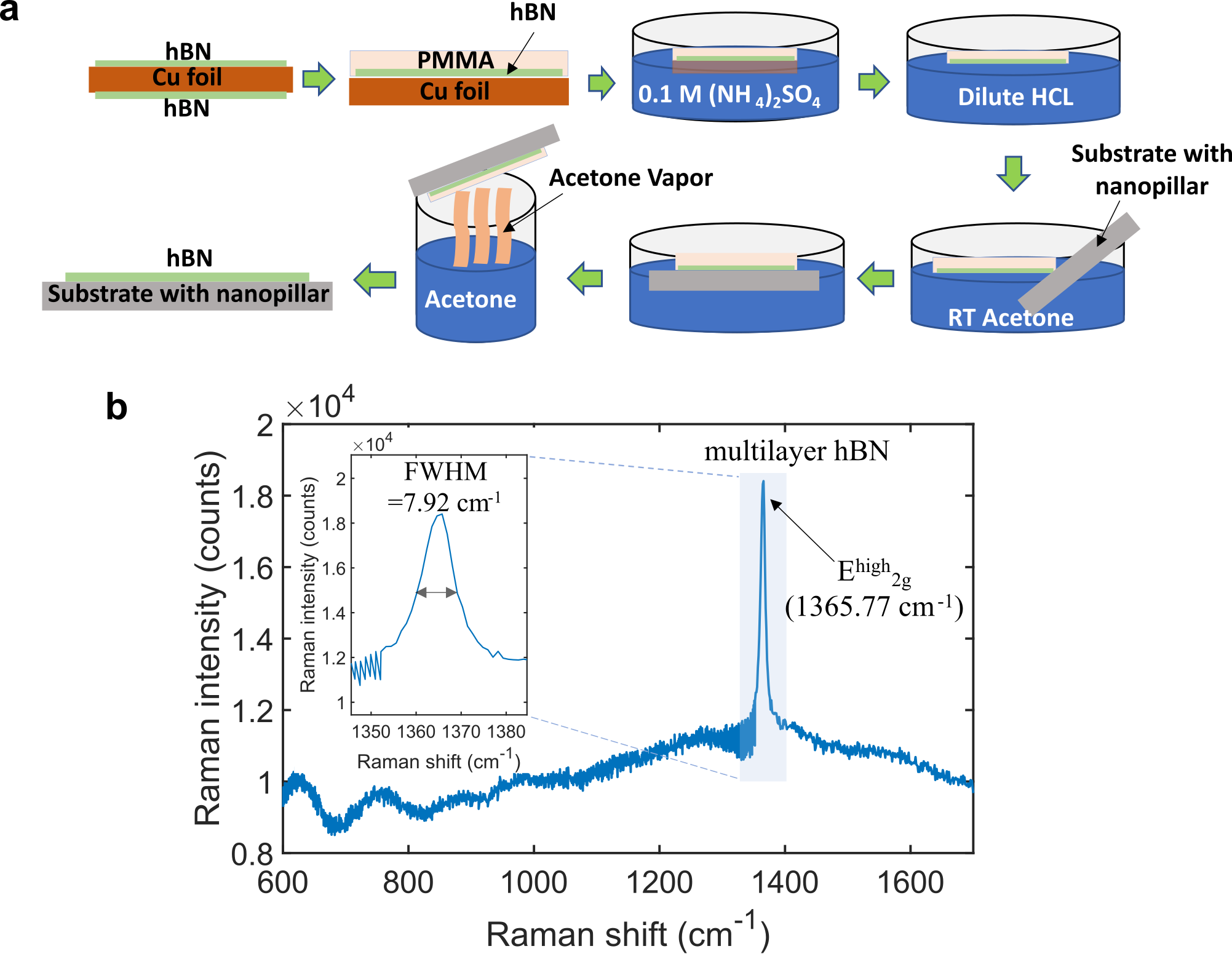}
\caption{Wet-transfer method and Raman measurement of a multilayered hBN film. (a) A step-by-step approach for transferring multilayer hBN film using a wet transfer method. (b) The characteristic hBN Raman peak ($\mathrm{E_{2g}}$) is recorded at 1,366~cm$^{-1}$. The inset shows the zoomed-in detail of the sharp spectral peak with an FWHM of 8~cm\textsuperscript{-1}.}
\label{Fig. S7}
\end{figure}
\newpage


\subsection{Section S2. FDTD Simulation of hBN-draped PNR}
\begin{figure}[h]
\centering
\includegraphics[width=1\textwidth]{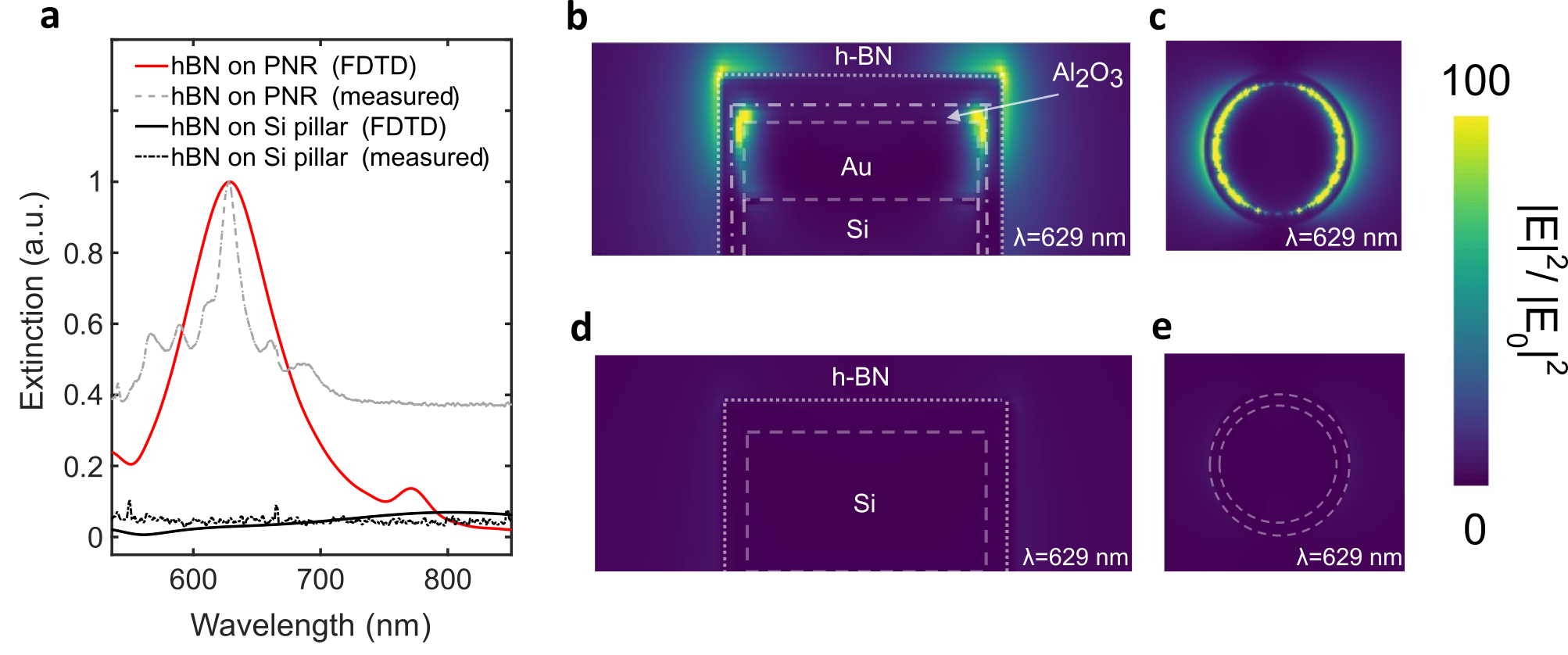}
\caption{Numerical simulation of a multilayer hBN draped on PNRs. (a) Simulated extinction spectra and measured PL intensity spectra are shown for two specific cases. Numerical simulation is performed with FDTD methodology. All the spectra are normalized. In the first case (red solid and grey dash-dot), a multilayer hBN (12~nm) film is draped along PNR. FDTD calculations (red solid) show a spectral peak appearing around 647~nm. Among the experimentally measured PL intensities, the one with the most spectral match has been presented in a grey dash-dot line. For the second case, the hBN film is draped along Si pillars. The corresponding FDTD and measured PL intensity spectra are represented with black solid and dash-dot lines, respectively. Here, hBN coupled to PNR shows a contrasting PL response over the case when hBN is draped along Si pillars. (b-e) Electric-field intensity distribution at the resonance wavelength of 647~nm for the cases when hBN is draped on (b and c) PNP and (d and e) only Si pillars. E-field intensity distribution shows a 100-fold enhancement supported by the PNR along its edges, generating strong plasmonic hot spots. On the other hand, when hBN is draped on Si pillars (d and e), the E-field intensity distribution shows no field enhancement.}
\label{Fig. S8}
\end{figure}
\newpage


\subsection{Section S3. Extraction of FOMs from SPE photoluminescence}
\begin{figure}[h]
\centering
\includegraphics[width=1\textwidth]{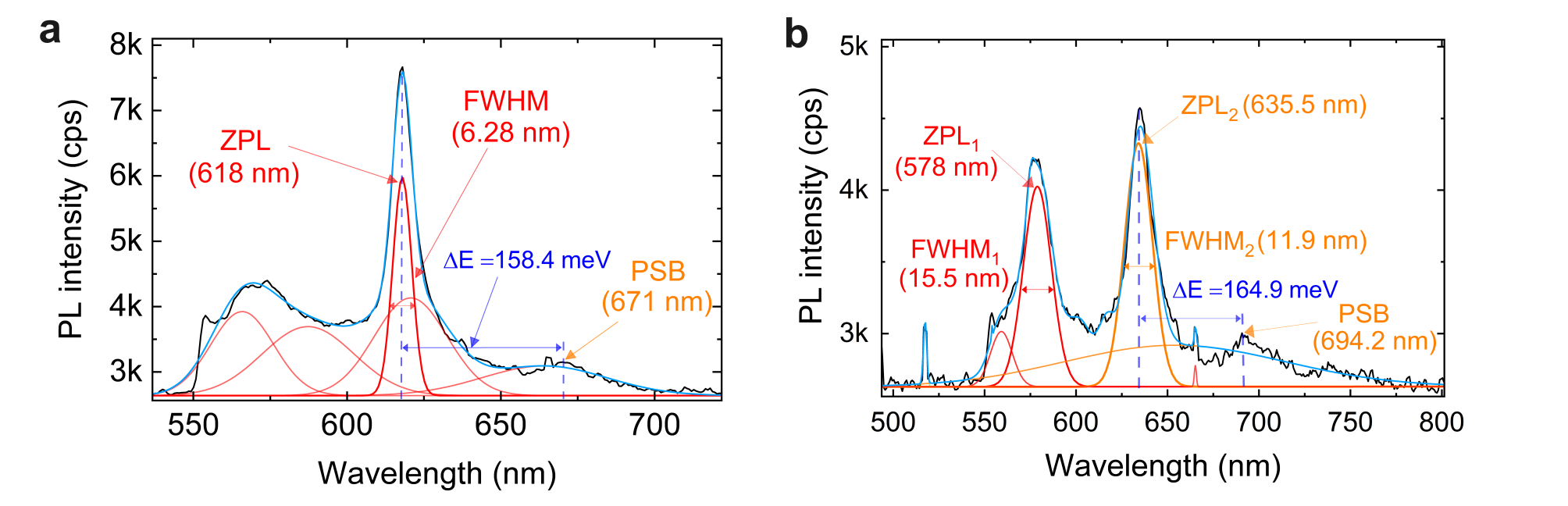}
\caption{Extraction of the figure of merits (FOMs) from SPE PL spectra (a and b) Two representative PL spectra of SPEs (with $g^{(2)}(0)<0.5$) recorded from different scans at room temperature. For the determination of PL transitions, the recorded PL spectra have been fitted with Gaussian lines to estimate ZPL and PSB peaks and their corresponding FWHMs.}
\label{Fig. S3}
\end{figure} 

The quality factor (Q) can be expressed as the ratio of the ZPL emission wavelength ($\mathrm{\lambda_p}$) to the relevant full width at half-maximum (FWHM). 
\[Q = \frac{\mathrm{\lambda_p}}{FWHM}\]
In Figure S3a, it shows the spectrum has a ZPL peak ($\mathrm{\lambda_p}$) at 618 nm. The corresponding FWHM is measured as 6.28 nm. So, the calculated quality factor (Q) is 98.3.
\newpage


\subsection{Section S4. Detailed analysis of SPE photoluminescence }
\begin{figure}[h]
\centering
\includegraphics[width=0.72\textwidth]{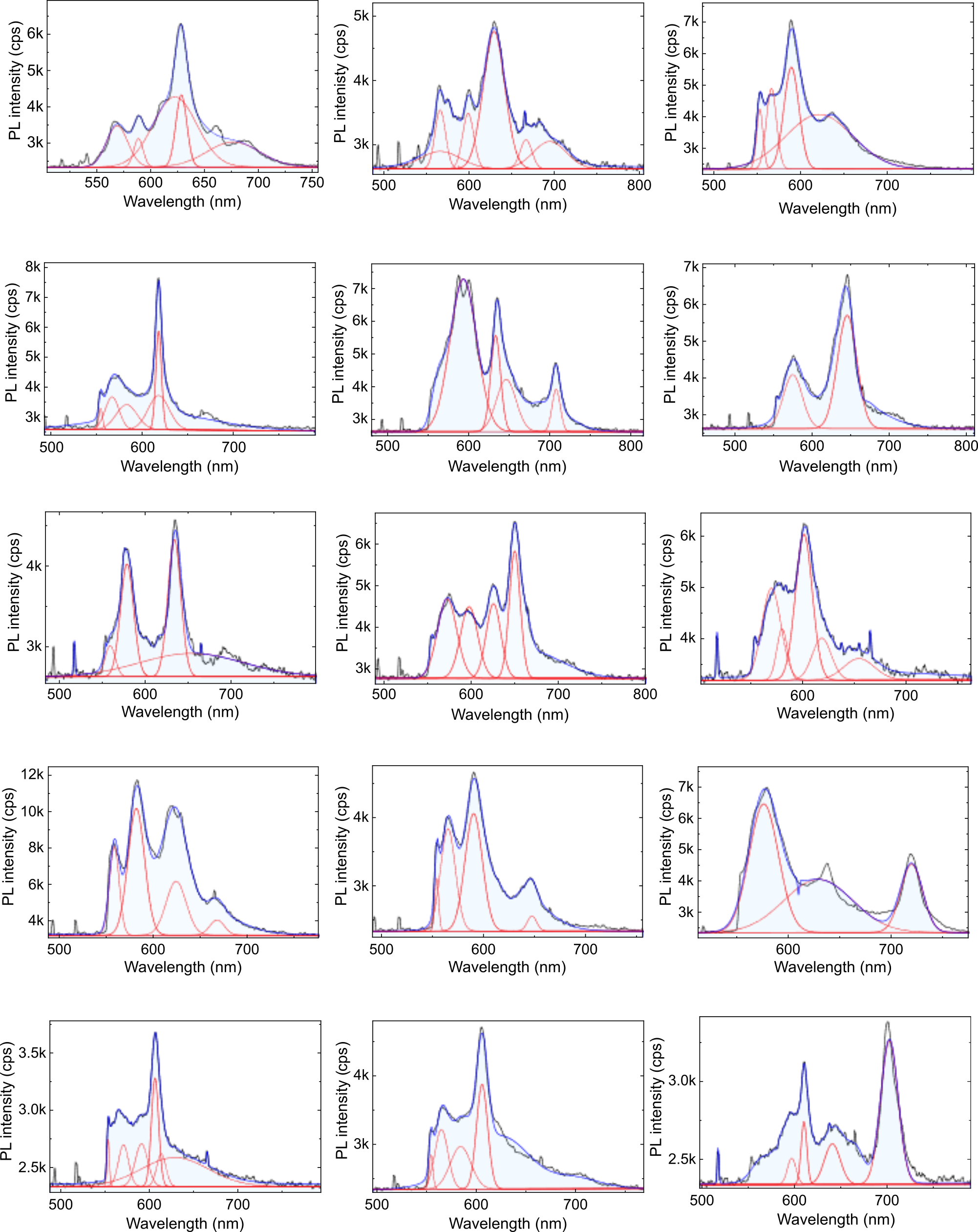}
\caption{Representative photoluminescence (PL) spectra of 15 out of the total 24 SPEs (with $g^{(2)}(0)<0.5$) recorded from different scans at room temperature. For the determination of PL transitions, the recorded spectra have been fitted with Gaussian lines to estimate the transition peaks and their corresponding FWHMs.}
\label{Fig. S1}
\end{figure} 
\newpage


\subsection{Section S5. Second-order autocorrelation $g^{(2)}(\tau)$ measurements}
\begin{figure}[h]
\centering
\includegraphics[width=0.67\textwidth]{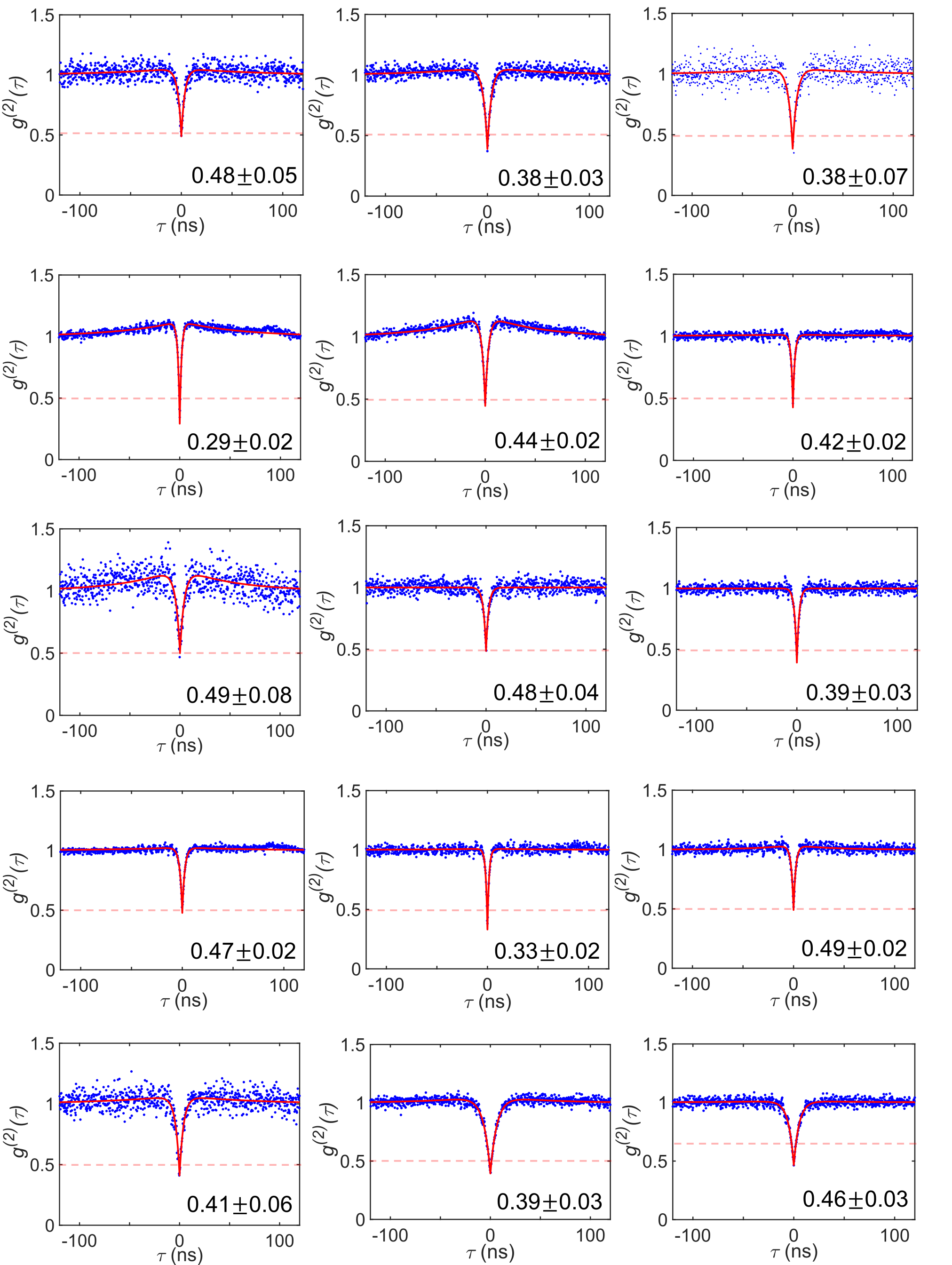}
\caption{Second-order autocorrelation $g^{(2)}(\tau)$ measurements for the corresponding 15 room-temperature SPEs represented in Supplementary Fig. S1. Here, solid traces represent theoretical fits to the experimental data using a three-level model of $g^{(2)}(\tau) = A+B\exp(-t/\tau_1) + C\exp(-t/\tau_2)$. Here, $\mathrm{\tau_1}$ and $\mathrm{\tau_2}$ represent the excited and metastable states relaxation times, respectively.}
\label{Fig. S2}
\end{figure} 
\newpage


\subsection{Section S6. Analysis of 24 recorded SPE photoluminescence}
\begin{figure}[h]
\centering
\includegraphics[width=1\textwidth]{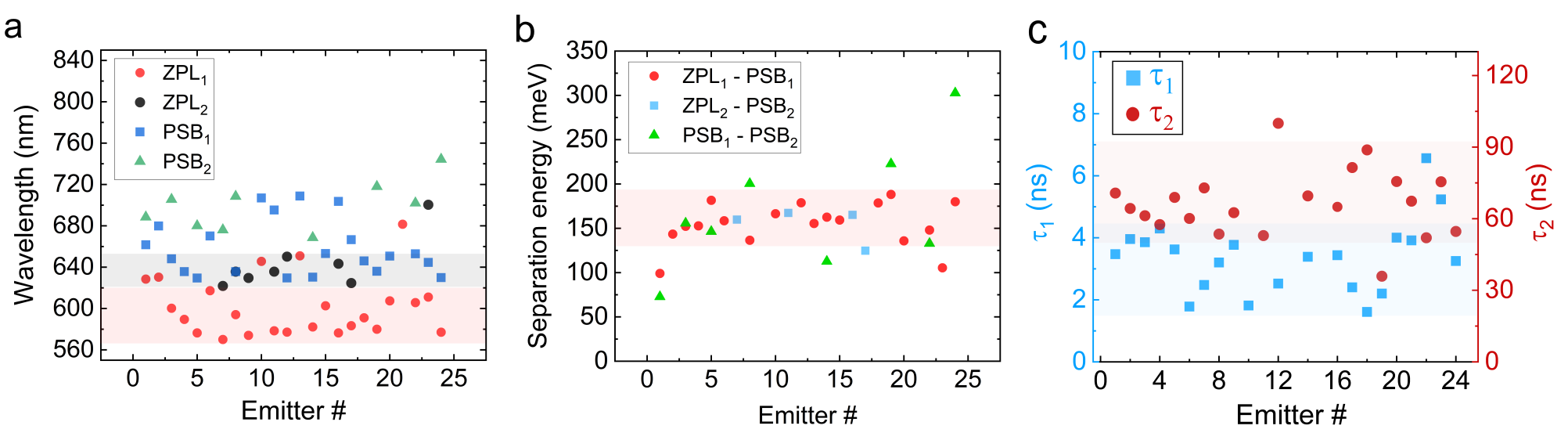}
\caption{Analysis of 24 SPE PL spectra (a) Wavelength distribution of PL spectral peaks captured from Gaussian fitting of 24 recorded room-temperature SPEs. The corresponding zero-phonon lines ($\mathrm{ZPL_1}$ and $\mathrm{ZPL_2}$), first phonon sideband ($\mathrm{PSB_1}$), and second phonon sideband ($\mathrm{PSB_2}$) peak positions are represented with red circles, black circles, blue squares, and green triangles, respectively. Statistics of ZPL distribution show that the emission wavelength responsible for $\mathrm{ZPL_1}$ and $\mathrm{ZPL_2}$ mostly remains between ~ 575 nm (2.15 eV) to ~ 610 nm (2.03 eV) (red shaded area) and ~607 nm (2.04 eV) to ~650 nm (1.91 eV) (grey shaded area), respectively. (b) Distribution of energy separation of peaks (i) from $\mathrm{ZPL_1}$ to $\mathrm{PSB_1}$ (with red circles), (ii) from $\mathrm{ZPL_2}$ to $\mathrm{PSB_2}$ (with blue squares), and (ii) from $\mathrm{PSB_1}$ to $\mathrm{PSB_2}$ (with green triangles), respectively. The energy separation from $\mathrm{ZPL_1}$ to $\mathrm{PSB_1}$ shows that the first phonon side band is redshifted from the $\mathrm{ZPL_1}$ by ~158±20 meV (red shaded area), approximately with an average energy of 154.59 meV. While the energy separation between $\mathrm{PSB_1}$ to $\mathrm{PSB_2}$ shows no obvious correlation. (c) The lifetimes of (i) excited states $\mathrm{\tau_1}$ and (ii) metastable states $\mathrm{\tau_2}$ are extracted from the SPE’s respective $g^{(2)}(\tau)$ measurements by fitting into a three-level model of $g^{(2)}(\tau) = A+B\exp(-t/\tau_1) + C\exp(-t/\tau_2)$. They are represented with square and circular markers, respectively.}
\label{Fig. S4}
\end{figure}
\newpage


\subsection{Section S7. Structural and optical characterization of hBN on PNRs}
\begin{figure} [h]
\centering
\includegraphics[width=0.9\textwidth]{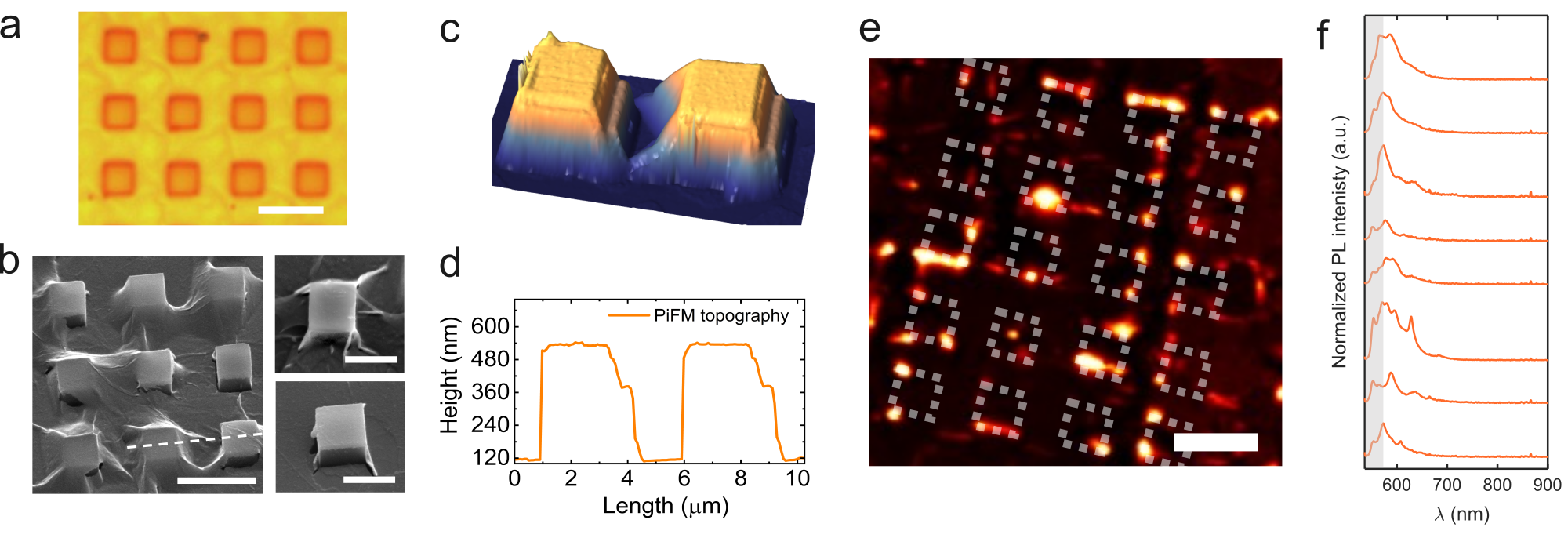}

\caption{Structural and optical characterization of hBN transferred on PNRs. (a) Optical image of hBN draped on PNRs. The scale bar represents 5 $\mu$m. (b) Scanning electron microscopy images of a 3 x 3 array of PNRs draped with hBN. The scale bar represents 5 $\mu$m. The dashed line shows a representative scanned section for topography extraction (left column). Single PNPs draped with hBN film. The mechanical draping profile varied across sites due to the hBN wet-transfer process onto the target device. We attribute these changes in the draping profile to have originated from the local strain introduced during hand-scooping the hBN onto the target device. Some PNRs are showing ‘tent-pole’ draping, while others represent hBN draping 'along and around' PNRs. Scale bars represent 2 $\mu$m (right column). (c) Three-dimensional (3D) and (d) Two-dimensional (2D) topography extracted photo-induced force microscopy (PiFM) measurement. (e) PL confocal mapping of a 5 x 4 array of PNRs draped with hBN. The scale bar represents 5 $\mu$m. (f) PL spectra recorded across different pillar sites. All the spectra are normalized without any background correction and vertically separated. The gray box indicates the 575 nm LP filter used while measuring the $g^{(2)}(\tau)$ measurements.} 
\label{Fig. S7.A.}
\end{figure}
\newpage


\begin{figure} [h]
\centering
\includegraphics[width=0.9\textwidth]{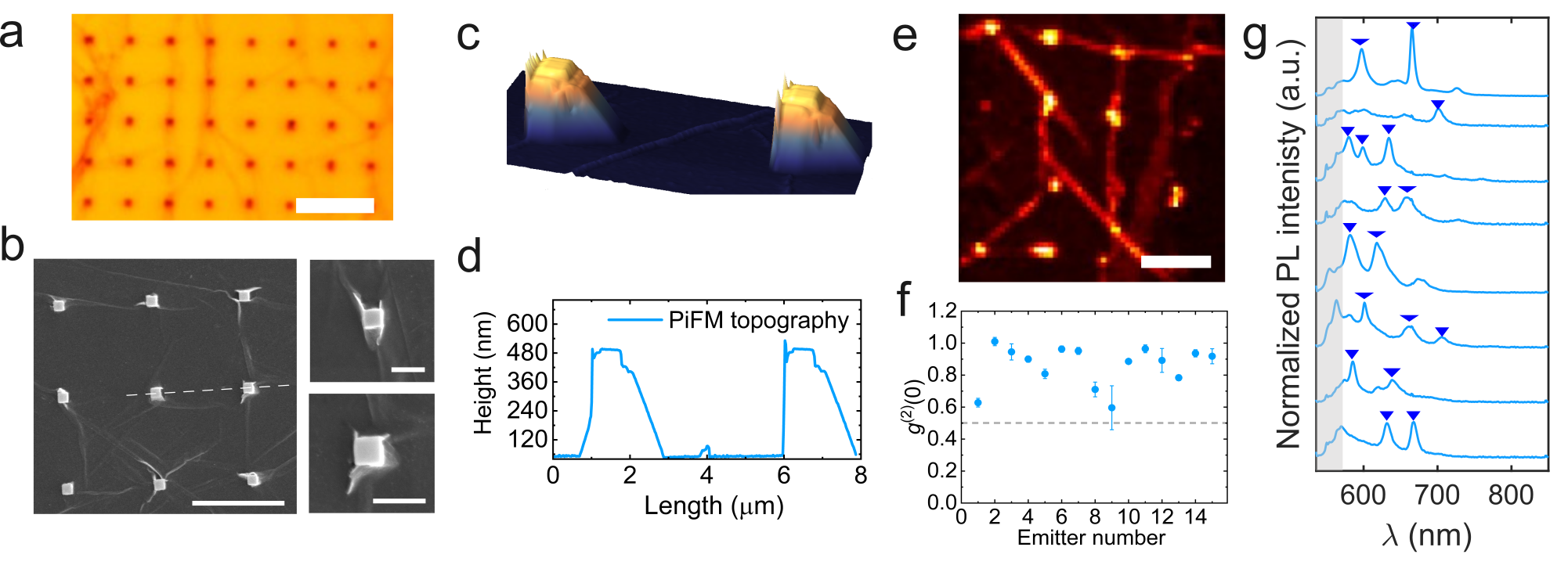}

\caption{Structural and optical characterization of hBN transferred on PNR. (a) Optical images of hBN transferred on PNRs. The scale bar represents 10 $\mu$m. (b) SEM images of a 3 x 3 array of PNRs draped with hBN. The scale bar represents 5 $\mu$m. The dashed line shows a representative scanned section for topography extraction (left column). Single PNRs draped with hBN. The mechanical draping profile varied across PNR sites. Some sites showed ‘tent-pole’ draping, while others have draped around PNRs. We attribute the change in the draping profile to have originated from the local strain introduced during hand-scooping hBN through wet transfer. The scale bar for the right column SEM pictures represents 1 $\mu$m. (c) 3D and (d) 2D topography extracted from PiFM measurement. (e) PL confocal mapping of a 4 x 4 array of PNRs with hBN. The scale bar represents 5 $\mu$m. (g) PL spectra recorded across different PNR sites. Spectra are not background corrected. Blue triangles indicate the multiple peaks behaving as few-photon emitters. The gray box indicates the 575 nm LP filter used while recording the $g^{(2)}(\tau)$ measurements. (f) Recorded $g^{(2)}(0)$ values plotted across different PNR sites draped with hBN.} 
\label{Fig. S7.B.}
\end{figure}
\newpage


\begin{figure} [h]
\centering
\includegraphics[width=0.9\textwidth]{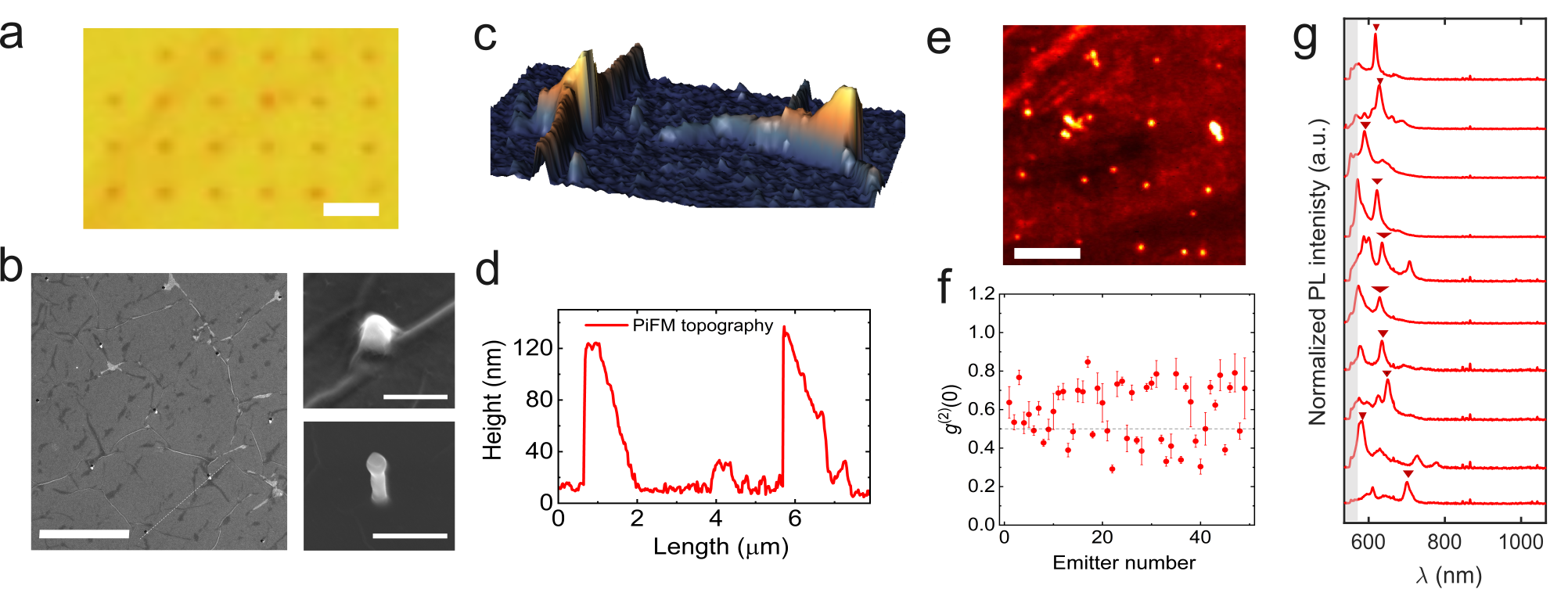}

\caption{Structural and optical characterization of hBN transferred on PNRs. (a) Optical images of hBN transferred on PNRs. The scale bar represents 5 $\mu$m. (b) SEM image of a 3 x 3 array of PNRs draped with hBN. The scale bar represents 5 um. The dashed line shows a representative scanned section for topography extraction. (left column). (c) 3D and (d) 2D topography extracted from PiFM measurement.  Single PNR draped with hBN. The mechanical draping profile varies across PNR sites. This variation in the draping profile could have originated from the local strain introduced during the hBN was being hand-scooped during the wet transfer. Some PNRs showed ‘tent-pole’ draping, while others represented hBN draping along and around PNRs. The scale bars for the SEM pictures represent 500 nm (right column). (e) PL confocal mapping of a 4 x 4 array of pillars draped with hBN. The scale bar represents 5 $\mu$m. (f) Recorded $g^{(2)}(0)$ values are plotted across different PNR sites with hBN. The emitters with a $g^{(2)}(0)<0.5$ can be considered as single photon emitters. (g) PL spectra recorded across different PNR sites. Spectra are not background corrected. Red triangles indicate the prominent emission peaks. The gray box indicates the 575 nm LP filter used while measuring the $g^{(2)}(\tau)$ measurements. }
\label{Fig. S7.C.}
\end{figure}
\newpage


\subsection{Section S8. Laser power-dependent second-order autocorrelation and fluorescence saturation measurements.}
\begin{figure} [h]
\centering
\includegraphics[width=0.95\textwidth] {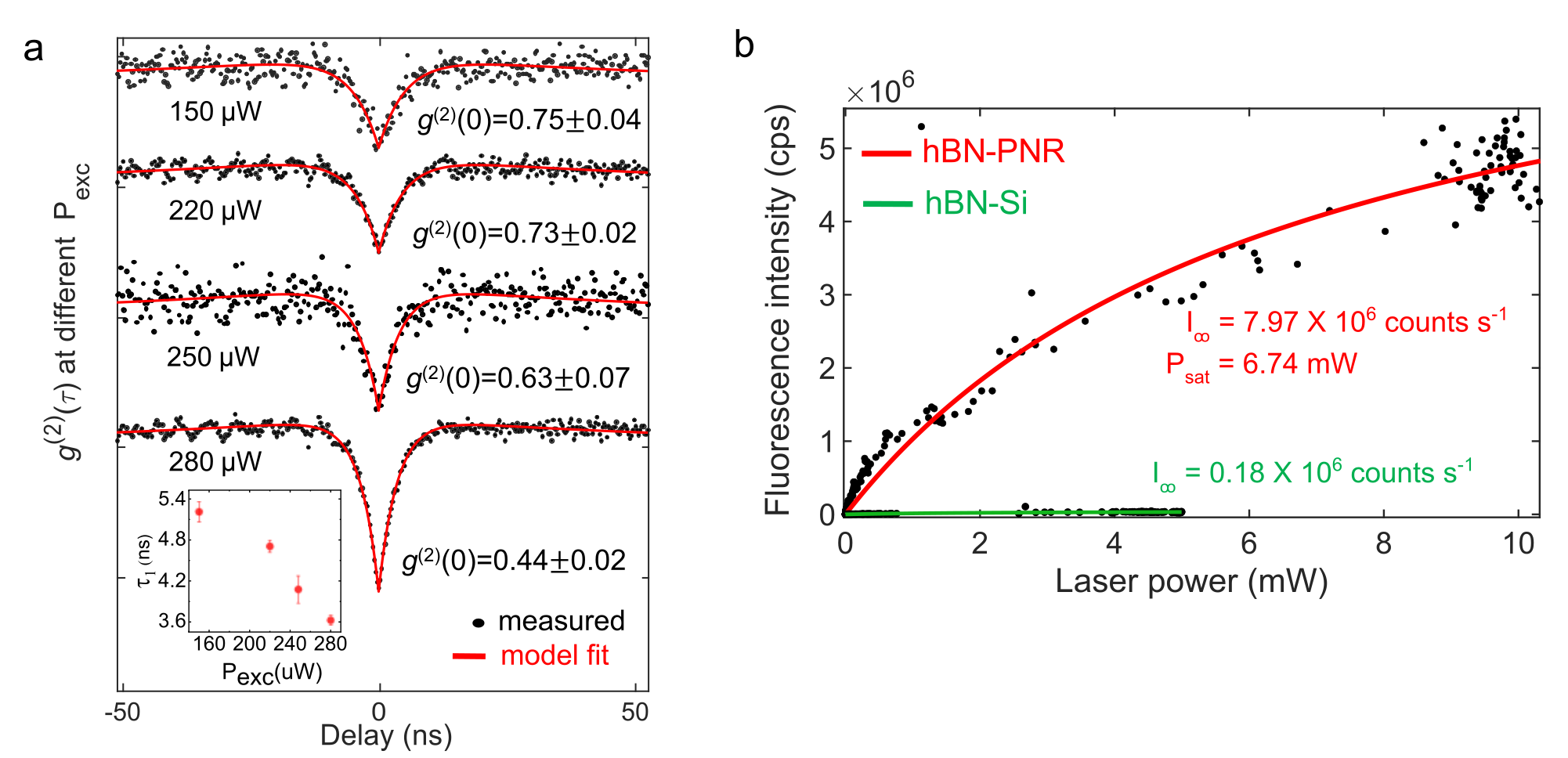}
\caption{Laser power-dependent second-order autocorrelation and fluorescence saturation measurements. (a) The SPE recombination dynamics is probed with a series of excitation-power-dependent second-order autocorrelation $g^{(2)}(\tau)$ measurements. Here, we observe a systematic decrease in the antibunching dip $g^{(2)}(0)$ value from 0.75(±0.04) to 0.44 (±0.02) as the incident excitation-power is increased in gradual steps from 150 ${\mu}$W to 280 ${\mu}$W. This gradual gain in photon purity is also accompanied by the narrowing of the antibunching peak which may be related to the decrease in their corresponding power-dependent emission lifetimes, $\mathrm{\tau_1}$ from 5.2 ns to 3.6 ns (as shown in the inset). Another feature in these $g^{(2)}(\tau)$ measurements is the presence of a photon bunching phenomenon along with the antibunching signature. Bunching behavior becomes most pronounced at the highest laser excitation power of the incident laser. Both these trends are in agreement with the previously reported works \cite{tran2016robust, senichev2021room}. (b) fluorescence saturation measurements recorded from hBN-alumina-PNP device configuration. Corresponding theoretical fits in solid lines are plotted using a three-level model. The saturated emission rates for this SPE has been measured as 7.97 Mcps with $\mathrm{P_{sat}}$ at 6.74 mW.}
\label{Fig. S6}
\end{figure}
\newpage


\subsection{Section S9. Schematic of measurement setups used for optical characterization}
\begin{figure} [h]
\centering

\includegraphics[width=1\textwidth]{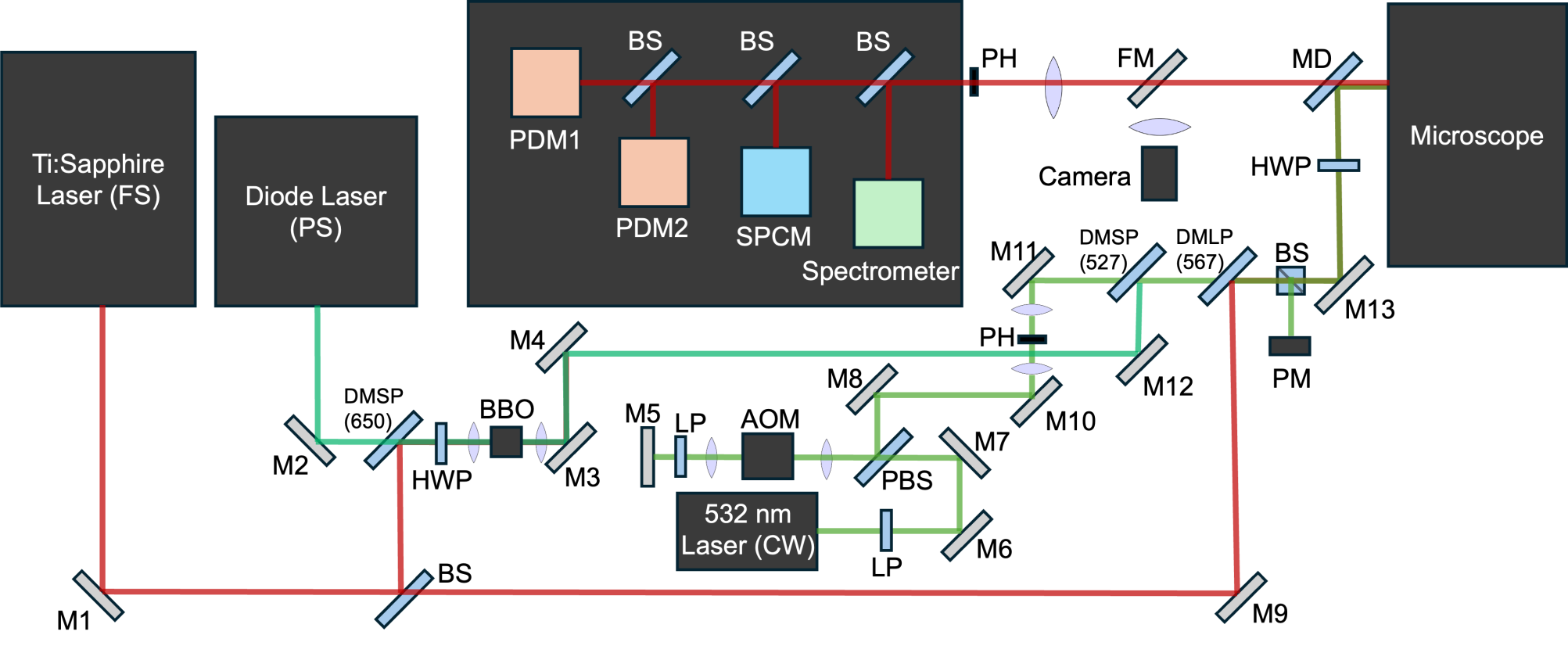}
\caption{Detailed schematic of the measurement setup. The lasers shown are a Mai Tai HP DeepSee Ti:Sapphire laser, Becker \& Hickl Picosecond Diode Laser, and a 532 nm  RGB  photonics laser. The Mai Tai is tunable from 690 - 1040 nm and has a 150 fs pulse width and 80 MHz repetition rate. The picosecond laser has a 515 nm center wavelength, 40 ps pulse width, and 80 MHz repetition rate. The RBG Photonics laser is centered at 532 nm and operates in the CW regime. The M1, M2, ..., etc. labeled objects are silver mirrors. BBO stands for beta barium borate, which is used for second harmonic generation. AOM stands for acousto-optical modulator, which is a Gooch \& Housego 3350-199. LP stands for linear polarizer. PH stands for pinhole filter, here it is 100 $\mu$m in diameter. PM stands for power meter. BS and PBS stands for beam splitter and polarizing beam splitter, respectively. HWP stands for halfwave plate. DMSP, DMLP, and MD stand for dichroic mirror short pass, dichroic mirror long pass, and main dichroic, respectively. The timing for detectors and devices is done with a 9520 series pulse generator by Quantum Composer. The SPCM is a Single  Photon  Avalanche  Diode (SPCM-AQRH, Excelitas), it has a 69\%  quantum efficiency at 650~nm and the PDMs are by Micro-Photon Devices. The spectrometer is by Ocean Insight,  a  QE65000  visible-to-near infrared spectrometer. This figure is a simplified version of the optical setup reduced to include only those components relevant to this experiment.}
\label{Fig. S8}
\end{figure}
\newpage


\subsection{Section S10. ZPL and PSB spectral peak distribution}
\begin{figure}[h]
\centering
\includegraphics[width=0.95\textwidth]{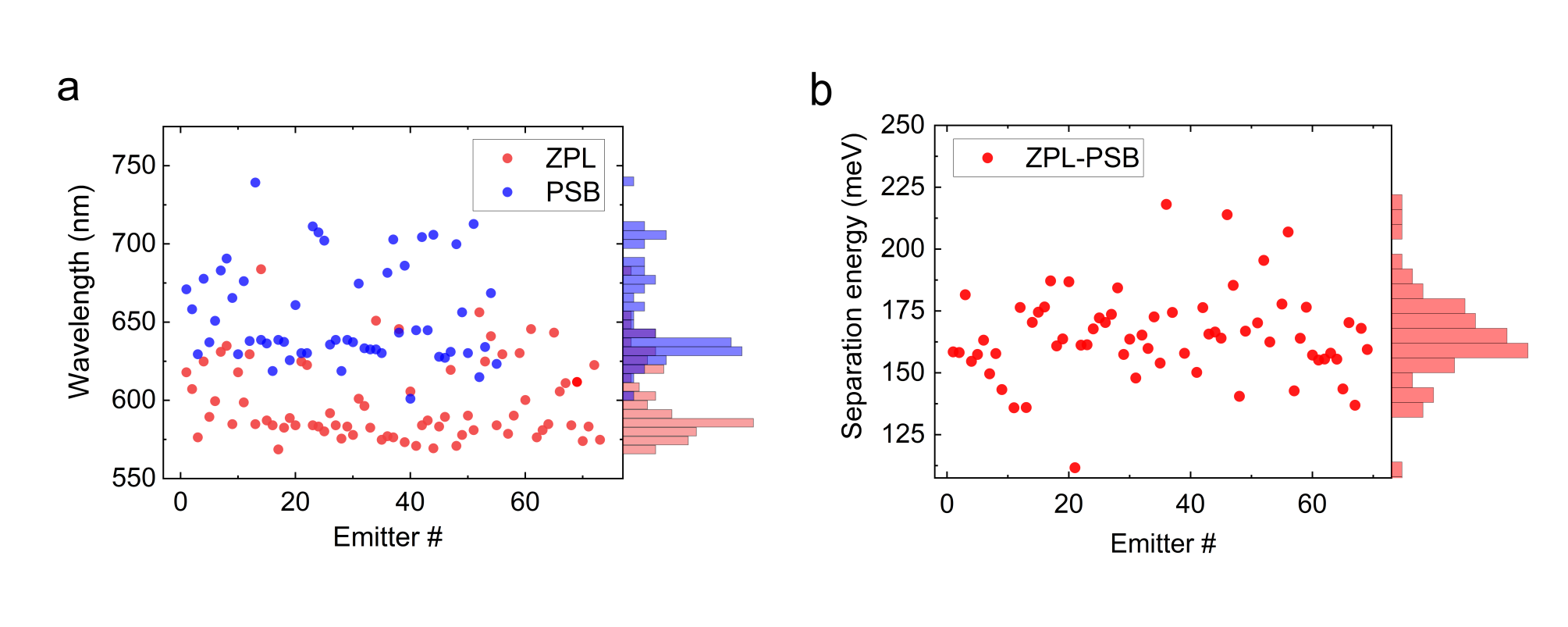}
\caption{Statistical distribution of PL spectral peaks recorded from a total of 73 emitters (a) Wavelength distribution of PL spectral peaks of 73 emitters recorded at room temperature. The corresponding zero-phonon line (ZPL) and phonon sideband (PSB) peak positions are represented with red and blue circles, respectively. Statistics of ZPL peaks show that they are distributed mostly in two regions. These regions span from 570 nm (2.17 eV) to 595 nm (2.08 eV) and from 607 nm (2.04 eV) to 645 nm (1.92 eV), respectively. (b) Distribution of energy separation between ZPL and PSB (red circles). The energy separation distribution shows that the PSB is redshifted from the ZPL with an overall average energy of 165 (± 10) meV. }
\label{Fig. S10}
\end{figure}
\newpage





\subsection{Section S11. DFT Calculation : $\mathrm{N_BV_N}$ on Gold}
\begin{figure} [h]
\centering
\includegraphics[width=0.9\textwidth]{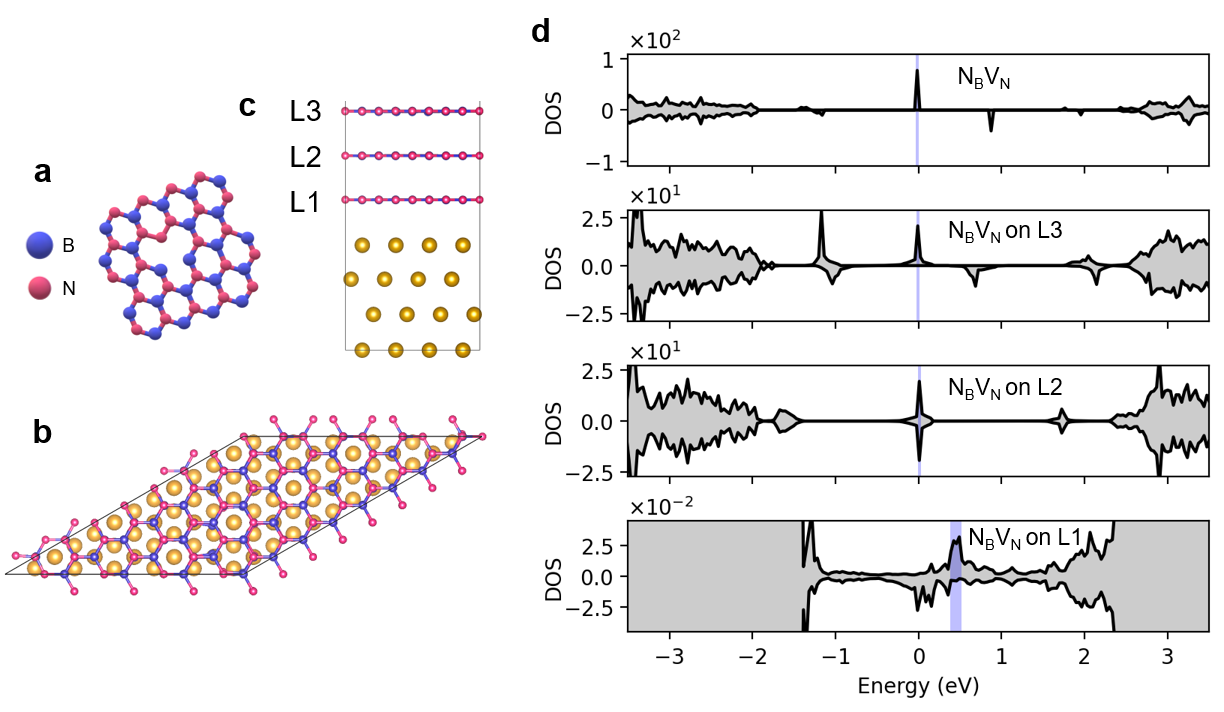}

\caption{Broadening of mid-gap state near hBN/Au interface for $\mathrm{N_BV_N}$ defects. (a) Depiction of $\mathrm{N_BV_N}$ defect (b) Top and (c) side vied of hBN/Au heterostructure used in the layer dependent defect calculations 
(d) Partial density of states (DOS) integrated over the three-layer hBN system in four configurations: with $\mathrm{N_BV_N}$ on the middle layer and no Au slab (top panel) and $\mathrm{N_BV_N}$ on each of the three layers in proximity to Au (bottom three panels). Full-width at half maximum (FWHM) of the mid-gap states shaded in blue shows broadening from the interaction with Au.}
\label{fig3}
\end{figure}
\newpage

\subsection{Section S12. DFT Calculation : $\mathrm{V_N}$ on $\mathrm{Al_2O_3}$}
\begin{figure} [h]
\centering
\includegraphics[width=0.9\textwidth]{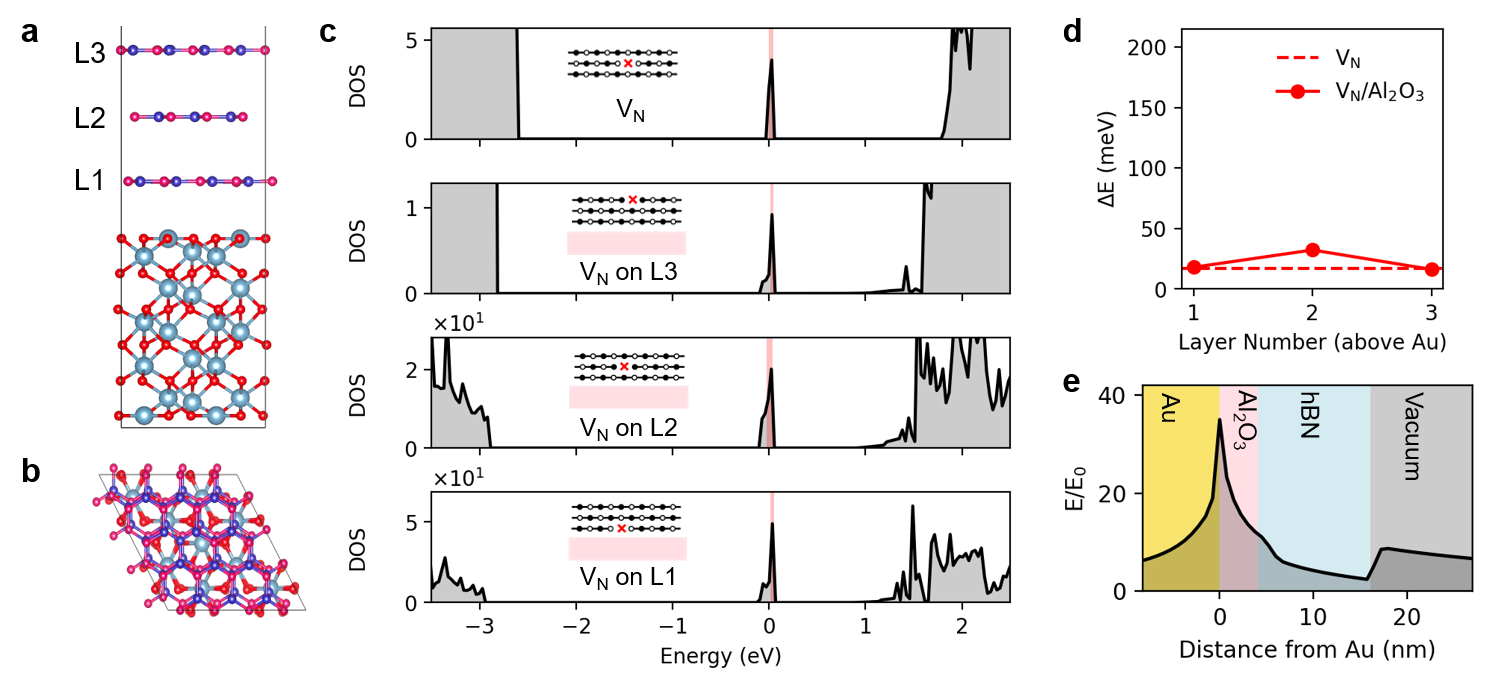}

\caption{Broadening of mid-gap state near hBN/$\mathrm{Al_2O_3}$ interface for $\mathrm{V_N}$ defects. (a) Side and (b) top view of the hBN/$\mathrm{Al_2O_3}$ heterostructure used in the layer-dependent defect calculations.  (c) Partial density of states (DOS) integrated over the three-layer hBN system in four configurations: with $\mathrm{V_N}$ on the middle layer and no $\mathrm{Al_2O_3}$ slab (top panel) and $\mathrm{V_N}$ on each of the three layers in proximity to $\mathrm{Al_2O_3}$ (bottom three panels). Full-width at half maximum (FWHM) of the mid-gap states shaded in red shows little broadening from the $\mathrm{Al_2O_3}$. (d) FWHM of the $\mathrm{V_N}$  mid-gap states as a function of layer number from the $\mathrm{Al_2O_3}$ surface. The dashed red line corresponds to defected hBN structure in the absence of $\mathrm{Al_2O_3}$. The same y-axis range from the main text is used here to illustrate a relative independence of the DOS on proximity to $\mathrm{Al_2O_3}$.  (e) Plasmonic enhancement of electric field versus distance from Au, obtained from Lumerical FDTD results for a PNP with an alumina spacer. The data presented corresponds to a horizontal cut positioned in the vicinity of the maximum field enhancement.}
\label{fig3}
\end{figure}

\end{suppinfo}


\end{document}